\documentclass{aa}  
\usepackage{graphicx}
\usepackage{amsmath, amssymb}
\usepackage{subfig}
\usepackage{booktabs}
\usepackage{natbib}
\usepackage{url}
\usepackage{hyperref}
\usepackage{longtable}
\usepackage{tablefootnote}
\usepackage{txfonts}
\begin{document}

   \title{Fast optical spectroscopic observations of PSR J1023+0038 over one orbital period}

   \author{M. M. Messa
          \inst{1}\inst{,2}, M. C. Baglio\inst{2}, P. D'Avanzo\inst{2}, G. Illiano\inst{2}, F.~Coti~Zelati\inst{3,4,2}, K. Alabarta\inst{2}, D. de Martino\inst{5}, Y.-D. Hu\inst{6}, A. Miraval Zanon\inst{7}, A. Reguitti\inst{8}\inst{,2} and S. Campana\inst{2}
          }

  \institute{
   Università degli Studi di Milano, Dipartimento di Fisica, Via Celoria 16, 20133 Milano (MI), Italy
         \and
             INAF, Osservatorio Astronomico di Brera, Via E. Bianchi 46, 23807 Merate (LC), Italy
        \and
        Institute of Space Sciences (ICE, CSIC), Campus UAB, Carrer de Can Magrans s/n, E-08193, Barcelona, Spain
        \and
        Institut d'Estudis Espacials de Catalunya (IEEC), E-08034 Barcelona, Spain
        \and
        INAF-Osservatorio Astronomico di Capodimonte, Salita Moiariello 16, I-80131 Naples, Italy
        \and 
        Guangxi Key Laboratory for Relativistic Astrophysics, School of Physical Science and Technology, Guangxi University, Nanning 530004, China
        \and
        ASI - Agenzia Spaziale Italiana, Via del Politecnico snc, I-00133 Rome, Italy
        \and
        INAF – Osservatorio Astronomico di Padova, Vicolo dell’Osservatorio 5, I-35122 Padova, Italy}
    
   \date{Received 28 April, 2026; accepted 25 July, 2026}

\titlerunning{Fast optical spectroscopy of J1023}
\authorrunning{M. M. Messa et al.}
 
  \abstract
   {Transitional millisecond pulsars (tMSPs) are neutron-star binaries that switch between rotation-powered and accretion-powered states, providing a key link between low-mass X-ray binaries and millisecond radio pulsars. In their sub-luminous disc state, these systems exhibit complex variability whose origin is still debated. We present high-time-resolution optical spectroscopic observations of the tMSP PSR J1023+0038 obtained during its sub-luminous disc state. Our dataset covers for the first time a full orbital cycle at minute-timescale cadence. We detect significant variability in the main properties of the optical emission lines, including the equivalent width (EW) and full width at half maximum (FWHM), on timescales of minutes. A comparison between the temporal evolution of these quantities reveals indications of correlated behaviour, with some FWHM minima coinciding with decreases in the EW. This may point to episodes of matter ejection from the inner regions of the accretion disc, possibly associated with switches to low modes. The Doppler tomography of the H$\alpha$ and H$\beta$ emission lines suggests the presence of asymmetric emission structures, consistent with a scenario in which part of the accreting material is expelled from the system. In addition, the optical continuum shows variability consistent with a possible orbital modulation associated with the irradiated companion star, although its characterisation is limited by the observing conditions. Our results provide new constraints on the short-timescale behaviour of tMSPs in the sub-luminous disc state and support scenarios in which accretion and outflow processes coexist. Further multiwavelength observations, particularly including simultaneous X-ray coverage, will be crucial to establish a direct link between the observed optical variability and the high/low mode switches.}


\maketitle
%
\section{Introduction}

Radio millisecond pulsars (MSPs) are rapidly rotating neutron stars that emit coherent radio pulsations at the spin period and have been spun up to millisecond periods through sustained mass transfer in low-mass X-ray binaries (LMXBs; \citealt{Alpar1982}; \citealt{Radhakrishnam&Srinivasan}). One of the most important pieces of evidence supporting this scenario, known as the recycling scenario (\citealt{Alpar1982}) for MSPs, came with the discovery of transitional millisecond pulsars (tMSPs), i.e. systems that alternate between long-lived rotation-powered and accretion-powered states (lasting years), with transitions occurring on timescales of days (\citealt{Campana&Disalvo2018}; \citealt{Papitto&DeMartino2022}). The first and best-studied example of a tMSP is PSR J1023+0038 (hereafter J1023), which is used as a prototype for this class of objects. Alongside J1023, other confirmed tMSPs are PSR J1824–2452I (M28I; \citealt{Papitto_2013}) and XSS J12270–4859 (\citealt{Bassa2014}; \citealt{DeMartino2013}). Additional candidates such as RXS J154439.4–112820, CXOU J110926.4–650224, 4FGL J0407.7–5702, 3FGL J0427.9–6704, 4FGL  J0639.1-8009 and 4FGL J1824.2+1231 (e.g., \citealt{Bogdanov2016}; \citealt{CotiZelati2019}; \citealt{Miller2020}; \citealt{Kyer2025}) are being actively investigated. Nevertheless, J1023 remains the most extensively observed and best-characterised system, offering unparalleled insight into the dynamics of transitional pulsars.

Initially identified as a radio source in the FIRST (Faint Images of the Radio Sky at Twenty-cm) survey (\citealt{Bond2003}), J1023 was classified as a cataclysmic variable in 2001 due to rapid optical flickering and a blue continuum with double-peaked emission lines, indicative of an accretion disc (\citealt{Szkody2003}). However, subsequent optical observations in 2003 and 2004 revealed the disappearance of flickering and emission lines, replaced by absorption features from a late-type G5 companion (\citealt{Woudt2004}; \citealt{Homer2006}; \citealt{Thorstensen_2005}). Spectroscopic studies found a 4.75-hour orbital period \citep{Thorstensen_2005,Archibald2009}. These findings suggested that J1023 was not a cataclysmic variable, but rather a quiescent neutron star X-ray binary. The identification of 1.69-ms radio pulsations (\citealt{Archibald2009}) confirmed it as a recycled MSP, revealing a system capable of switching between accretion and rotation-powered states. In 2013, J1023 underwent a dramatic transformation. Its radio pulsations vanished, and a sudden five-to-ten-fold increase in the X-ray and gamma-ray flux was observed, along with brightening in the UV and optical by 1–2 magnitudes \citep{Stappers2014,Patruno2014}. Optical spectroscopy revealed the reappearance of double-peaked emission lines, signalling the formation of a new accretion disc \citep{CotiZelati2014}. Since then, J1023 has remained in this sub-luminous accretion disc state, with a persistent X-ray luminosity of $\sim 7 \times 10^{33} \ $erg \ s$^{-1}$ in the 0.3–80 keV range \citep{CotiZelati2018}, at a distance of 1.37 kpc \citep{Deller2012}. This is well below the typical luminosities of active LMXBs ($\sim 7 \times 10^{36} - 10^{38}\ $erg \ s$^{-1}$).

In this sub-luminous disc state, J1023 exhibits three distinct X-ray modes of emission: a high mode (L$_X$ $\sim 7 \times 10^{33} \ $erg \ s$^{-1}$, $\sim$ 70–80\% of the time), a low mode (L$_X$ $\sim 3 \times 10^{33} \ $erg \ s$^{-1}$, $\sim$20–30\%), and brief flares reaching $\gtrsim 10^{34}-10^{35} \ $erg \ s$^{-1}$ ($\sim$ 1–2\% of the time, \citealt{Linares2014}; \citealt{Bogdanov2015}; \citealt{Archibald2015}). Importantly, the spin down luminosity of J1023 is $\simeq$ 4.4 $\times 10^{34}$ \ erg \ s$^{-1}$ \citep{Archibald2013}, roughly six times brighter than its average X-ray luminosity. This suggests that the rotational kinetic energy of the pulsar remains a major contributor to the system’s energetics even in the sub-luminous disc state. Mode switches occur rapidly on $\sim$10–30 s timescales, with low modes typically lasting from several tens of seconds to a few minutes. Coherent pulsations at the neutron star spin period are detected in X-ray, UV, and optical bands during the high mode (and occasionally in flares), but vanish in the low mode \citep{Archibald2015,Papitto2019,Ambrosino2017,Zanon2022,Illiano2023}. Because this behaviour is difficult with accretion models, this is interpreted as evidence that the radio pulsar remains active even during the sub-luminous accretion disc state. During high modes, one possible scenario is that energy is dissipated at the interaction region between the relativistic pulsar wind and the innermost accretion flow, located just outside the light-cylinder radius ($\sim$80–100 km from the neutron star), producing synchrotron emission across multiple bands (\citealt{Papitto2019}; \citealt{Veledina2019}; \citealt{BaglioZelati2023}), which is thought to be responsible for the observed pulsed emission. However, recent results by \cite{Jadoliya2026} suggest an alternative picture in which, during the high mode, the accretion disc may extend inside the light-cylinder radius. In this scenario, the detection of a narrow Fe line attributed to reflection provides independent evidence for the presence of an inner disc component. It also interprets the spin pulses as accretion-induced onto the NS, challenging the shock interpretation. During low modes, the interaction point may move outward (\citealt{Papitto2019}), possibly as a consequence of the evacuation of the innermost accretion flow region \citep{BaglioZelati2023}, or the accretion flow may penetrate inside the light cylinder, leading to the activation of the propeller regime \citep{Veledina2019}. In either of these configurations, the pulsar wind flows around the disc instead of colliding with it, reducing dissipation and inhibiting pulsations.

The flaring mode, instead, may result from occasional increased thickness of the inner disc that intercepts the pulsar wind (\citealt{Papitto2019}; \citealt{Veledina2019}), but clear observational confirmation is still lacking. Recent observations have shown that X-ray flaring modes correspond to prominent optical and radio flares, possibly linked to an increased mass loading of the jet \citep{Baglio2025a}.
Rapid millimetre-band flares at the high-to-low mode switches have been interpreted as signatures of optically thin ejecta as the inner disc is expelled (\citealt{BaglioZelati2023}). J1023 also exhibits bright and variable radio continuum emission \citep{Deller2015} that is anti-correlated with the X-ray flux \citep{Bogdanov2018}: radio brightness increases during X-ray low modes, with the spectral index evolving from flat to inverted. This is consistent with the emission of optically thin ejecta on top of a compact, self-absorbed radio jet \citep{Bogdanov2018,BaglioZelati2023}. 

At UV, optical, and near infrared (NIR) wavelengths, the emission is dominated by the accretion disc and the irradiated companion star. However, the system also shows frequent flaring and flickering activity, attributed to a mix of reprocessed emission and synchrotron radiation from plasmoids within the accretion flow \citep{Shabaz2018,Kennedy2018,Papitto2018,Hakala2018,Baglio2019}. Time-resolved spectroscopy and Doppler tomography of the H$\alpha$ line have revealed changes in the disc structure that may suggest transient propeller behaviour or, more generally, matter ejection during flares \citep{Shahbaz2015,Hakala2018}. Recent strictly simultaneous optical and X-ray polarimetric studies have revealed a linear polarisation of $\sim$1\% in the optical and $\sim$12\% in soft X-rays (2-6 keV) during high modes, with consistent polarisation angles in the two bands \citep{BaglioZelati2023,Baglio2025b}. 
Multiband polarimetric and time-resolved observations continue to provide critical clues toward resolving the nature of the emission mechanisms.

In this paper, we report the results of the first high time-resolution optical spectroscopic campaign of the transitional millisecond pulsar J1023 covering an entire orbital period in order to study the possible variations of the main emission line properties across the entire orbital period. Using the Gran Telescopio Canarias (GTC), we obtained approximately one spectrum per minute, covering slightly more than a full orbital period of the system. An exploratory search sampling only 22\% of the orbital period has been reported in \cite{Messa2024}.\\
The paper is organised as follows. In Sect. \ref{Sec::Data_reduction}, we describe the observations and data reduction. The results and their discussion are presented in Sects. \ref{Sec::Results} and \ref{Par:discussion}, respectively. Finally, the conclusions are summarised in Sect. \ref{Sec::Conclusions}.

\section{Data reduction and analysis}\label{Sec::Data_reduction} 
The spectra of J1023 were obtained with the 10.4 m GTC at the Observatorio del Roque de los Muchachos, La Palma (Canary Islands, Spain). The OSIRIS (Optical System for Imaging and low-Intermediate-Resolution Integrated Spectroscopy; \citealt{Cepa2000}) instrument was used for long slit spectroscopy with a 1.5 arcsec slit width and the R1000B grism, reaching a resolution ($\lambda/\Delta \lambda$) of $\sim$ 1000 with a dispersion value of 2.12 $\AA$/pixels and spectral coverage between 3630 and 7500 $\AA$. Observations were carried out during the night between the 30 and 31 March, 2024 from 20:52 UT to 02:59 UT, covering $\sim1.3$ times the 4.75h orbital period of the binary, from phase 0.22 to 1.51, where orbital phase 0 corresponds to the inferior conjunction of the companion, when the non-irradiated face of the star is visible to the observer. The orbital phase values are based on the most recent ephemeris of \cite{Illiano2023}, adopting the ascending node epoch from the X-ray observations performed in 2022 with NICER. The corresponding orbital phases were computed using the \texttt{molly}\footnote{\url{https://cygnus.astro.warwick.ac.uk/phsaap/software/molly/html/INDEX.html} software package, which calculates the orbital phase from the heliocentric Julian date (HJD) of each spectrum.} Each spectrum was obtained with an exposure time of 20 s, yielding a total of 480 spectra with a cadence of approximately 50 s due to the CCD readout time. The reduction was performed using specialised pipelines such as the GTC/OSIRIS Data Reduction System (which uses Python-based tools like PypeIt\footnote{\url{https://pypeit.readthedocs.io/en/stable/}}), providing as final products spectra that are both wavelength and flux calibrated. The wavelength calibration is carried out using HgAr and Ne arc lamps, while the flux calibration is based on observations of the spectrophotometric standard star Hiltner 600. Due to the weather conditions of the night, all the spectra had to be corrected for slit losses based on the observed seeing (from 1.2 to 4.8 arcsec). The correction was made assuming a Gaussian point spread function (PSF) dominated by atmospheric seeing. For each exposure, the seeing FWHM was measured and used to compute the fraction of the total flux transmitted through the 1.5 arcsec slit. The transmitted fraction was calculated by integrating the Gaussian PSF across the slit width. Each spectrum was then corrected by dividing the observed flux by this fraction. Errors were propagated accordingly. Due to the lack of simultaneous X-ray coverage, Swift/XRT observations of J1023 acquired in March and April 2024 have also been analysed (results are reported in Appendix \ref{appendix:A}) to confirm unambiguously that the source was in the sub-luminous disc state at the time of our optical spectroscopic campaign. Indeed, the spectral features (emission lines) and average flux during the observation were characteristic of a sub-luminous state and refer to Sect. \ref{Sec::Results}.

\section{Results}\label{Sec::Results}
\subsection{Average spectrum}
\begin{figure}[h!]
   \centering
   \includegraphics[width=1\linewidth]{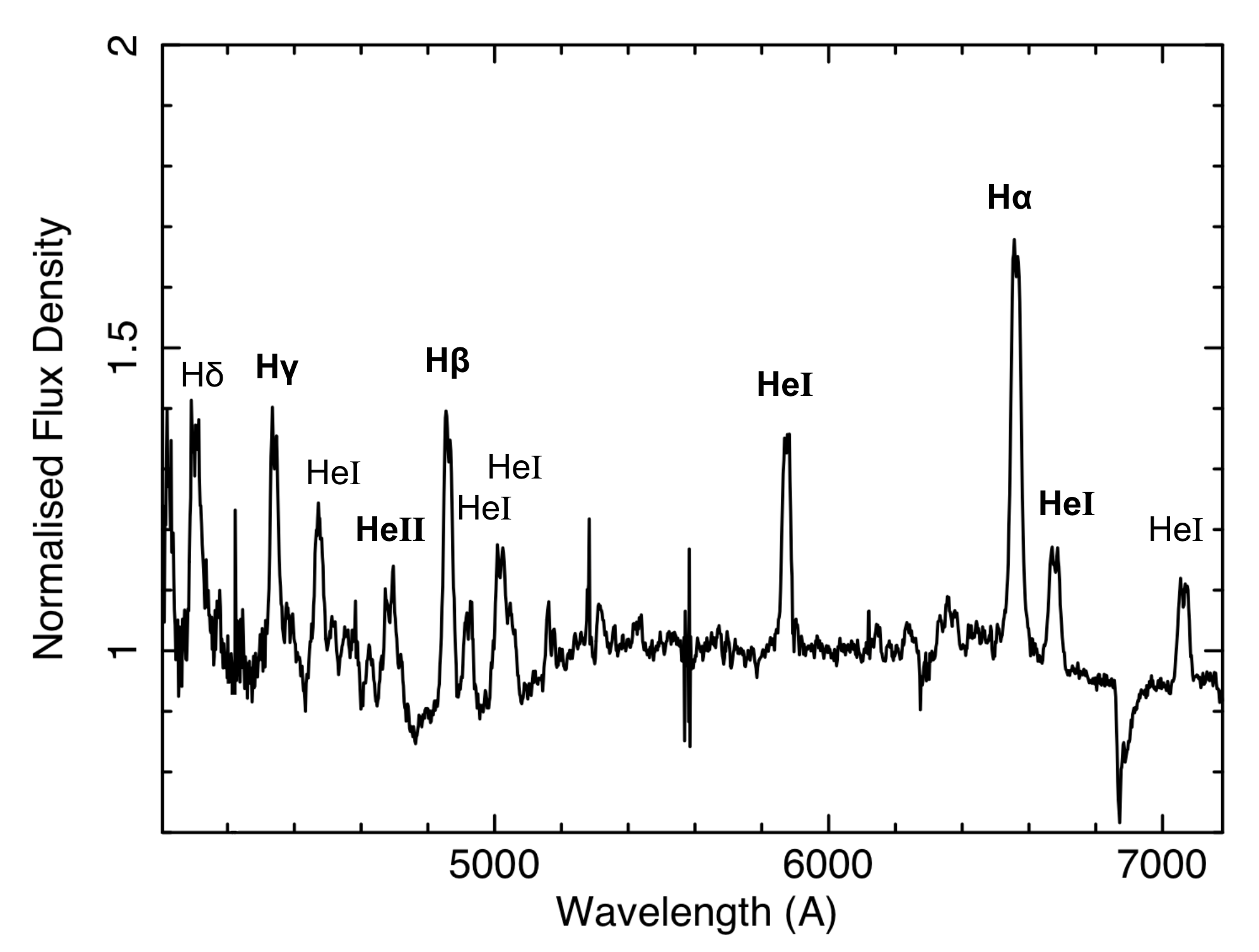}
      \caption{Average spectrum of J1023 normalised to the emission of the continuum. The emission lines studied in this work are highlighted in bold.
              }
         \label{fig:average}
   \end{figure}

Figure \ref{fig:average} shows the normalised and averaged (not phase corrected) spectrum of J1023 obtained in our campaign. We detect clear emission from H$\alpha$, H$\beta$, H$\gamma$, H$\delta$, He I ($\lambda$4472, 4921, 5016–5048, 5876, 6678, 7065 AA), He II at 4686 $\AA$ and from the Bowen blend. In agreement with previous studies \citep{CotiZelati2014,Linares2014,Hakala2018,Messa2024}, these lines (except for Bowen blend) exhibit a characteristic double-peaked profile, consistent with emission from an accretion disc, as expected from a tMSP in the sub-luminous disc state seen at moderate inclinations.

\noindent

\subsection{Continuum spectrum}
\begin{figure}[h!]
   \centering
   \includegraphics[width=1\linewidth]{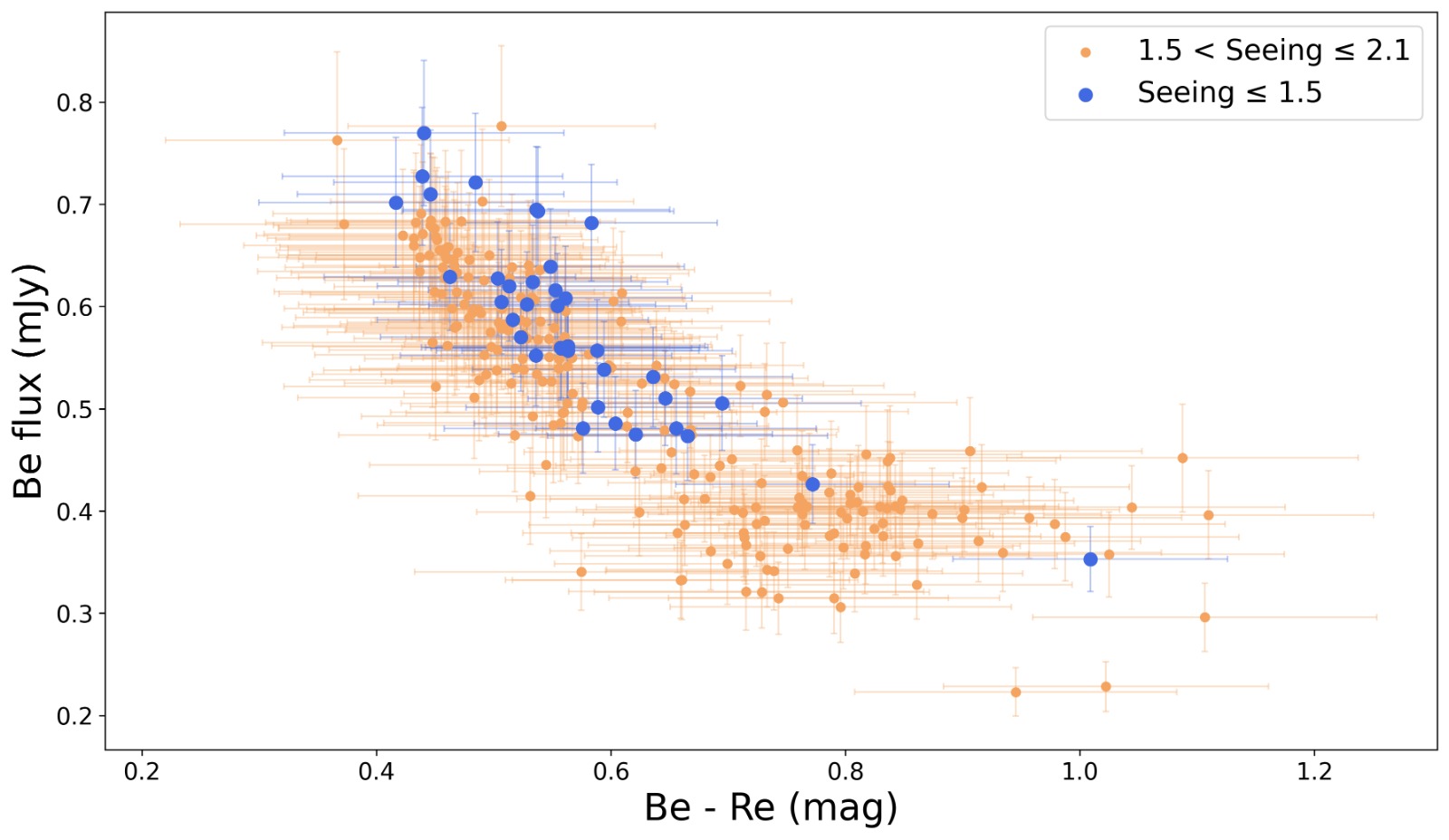}
      \caption{Colour-magnitude diagram for J1023 derived from Re and Be bands. The different colours of the points underline the different values of the seeing.  
              }
         \label{fig:color-magnitude}
   \end{figure}

\begin{figure}[h!]
   \centering
   \includegraphics[width=1\linewidth]{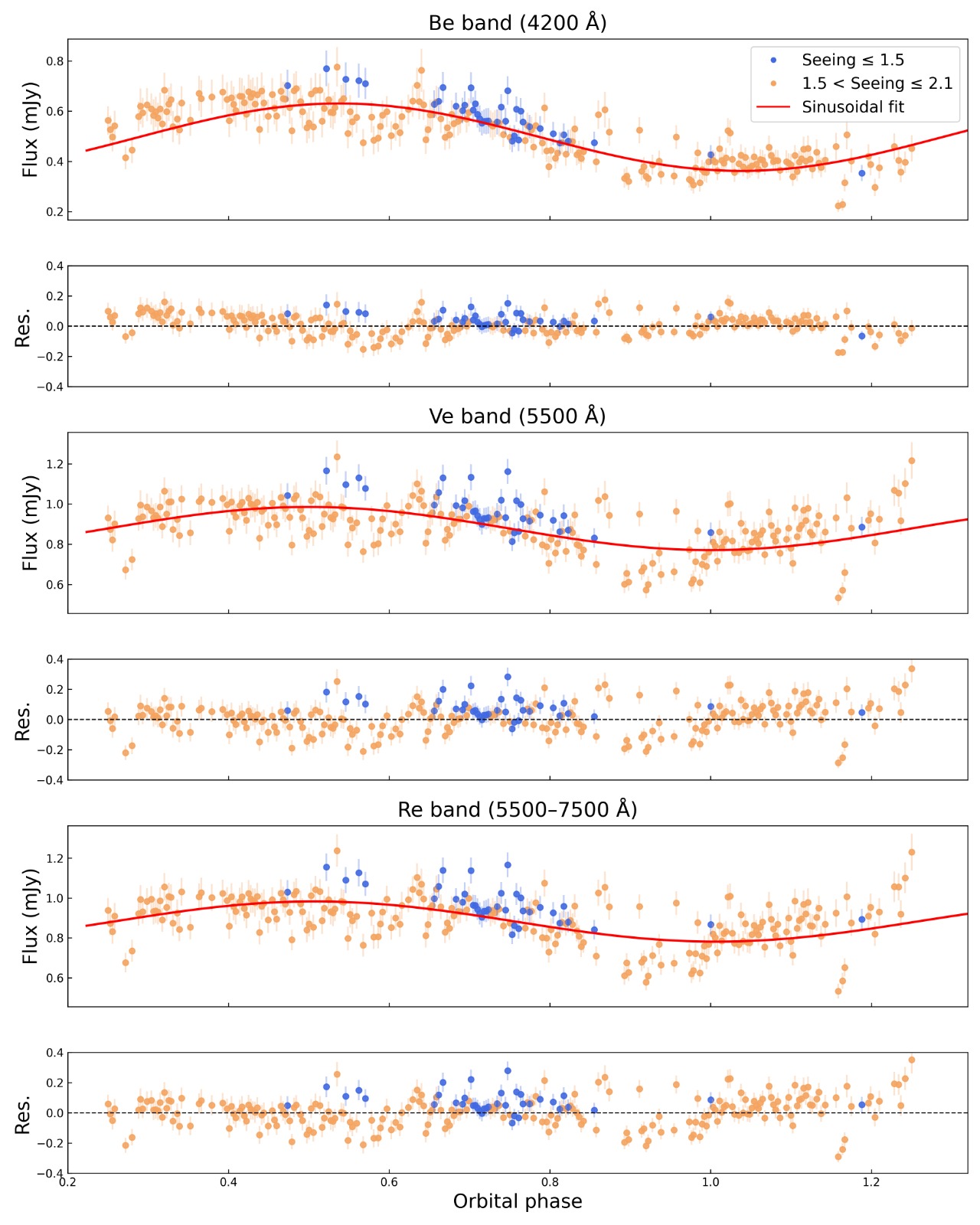}
      \caption{Orbital-phase dependence of the continuum flux in the Be, Ve, and Re bands. The different colours of the points underline the different values of the seeing. The red solid curves represent sinusoidal models (sine plus constant with period fixed to 1) fitted to the entire subset of data obtained with seeing $\leq$ 2.1" (blue and orange points). The bottom panels display the residuals of the fits.}
         \label{fig:continuum-fit}
   \end{figure}

To investigate the temporal evolution of the spectral properties, we first examined the behaviour of the optical continuum at different wavelengths. This analysis was carried out using the \texttt{molly} software, quantifying the continuum variability by selecting a wavelength interval. This was first for a wavelength interval approximately corresponding to the R-band region, from 5500 to 7500 $\AA$ (hereafter Re), while masking all emission-line features. The analysis was then repeated for narrower wavelength intervals including a region of width $\sim$ 120 $\AA$ centred at 4200 $\AA$ ($\sim$ B band, hereafter Be) and a region of width $\sim$ 600 $\AA$ around 5500 $\AA$ ($\sim$ V band, hereafter Ve). Thanks to this, it was possible to construct a colour–magnitude diagram by comparing the light curve centred at the Be band with that centred on the Re band, as a function of the flux calculated in the Be band. The result is shown in Fig. \ref{fig:color-magnitude}, which displays a decreasing trend, with the spectra becoming progressively redder (i.e. increasing Be-Re) towards lower Be intensities. Given that the colour can vary significantly depending on the seeing conditions at the time of the observation, this plot was produced by imposing two different seeing thresholds. The blue points represent spectra obtained under seeing conditions better than 1.5 arcsec (i.e. comparable to the slit width corresponding to a total of 28 spectra), while the orange points correspond to a more relaxed threshold of 2.1 arcsec. This value was adopted as a compromise between data reliability and statistical completeness of the sample, resulting in the exclusion of 225 spectra (approximately 47\% of the sample).\\
\noindent
To investigate the temporal evolution of the optical continuum, computed as described above, we analysed its dependence on the orbital phase to infer possible sinusoidal or ellipsoidal modulation known to be present \citep{Kennedy2018,Papitto2018}. Although slit-spectroscopic data are not ideally suited for this purpose, such an analysis provides a simultaneous diagnostic of the optical light-curve variability, allowing a direct comparison with the spectroscopic diagnostics presented in this work and helping to further characterise the behaviour of the accretion disc. Fig. \ref{fig:continuum-fit} shows the light curves in three different bands, Be, Ve, and Re, with different colours indicating the seeing conditions of the individual spectra, as described above. For each band, a fit consisting of a constant plus a sinusoidal component with the period fixed to the system orbital period was applied to the entire subset of spectra with seeing $\leq$ 2.1 arcsec (blue and orange points). The significance of the fit has been compared with a simple constant through an F-test, providing values of 1$\times$10$^{-16}$ for the Be, Ve, and Re bands, i.e. $>8\, \sigma$. The resulting modulation is consistent with a sinusoidal trend, peaking around orbital phase $\simeq$ 0.5 (i.e. the inferior conjunction of the neutron star when the observer sees both the irradiated face of the companion star and the accretion disc) and reaching a minimum near phase $\simeq$ 1. The best-fit sinusoidal model yields amplitudes of $0.134 \pm 0.004$ mJy, $0.107 \pm 0.005$ mJy, and $0.101 \pm 0.005$ mJy for the Be, Ve, and Re bands, respectively. The residuals of the fits are shown in the bottom panels. In conclusion, the combined analysis of Figs. \ref{fig:color-magnitude} and \ref{fig:continuum-fit} indicate that the continuum becomes bluer at orbital phases corresponding to the optical maximum.
\noindent

\subsection{Equivalent width}\label{Sec:EW}
\begin{figure*}[h!]
\centering
{\includegraphics[width=.315\textwidth]{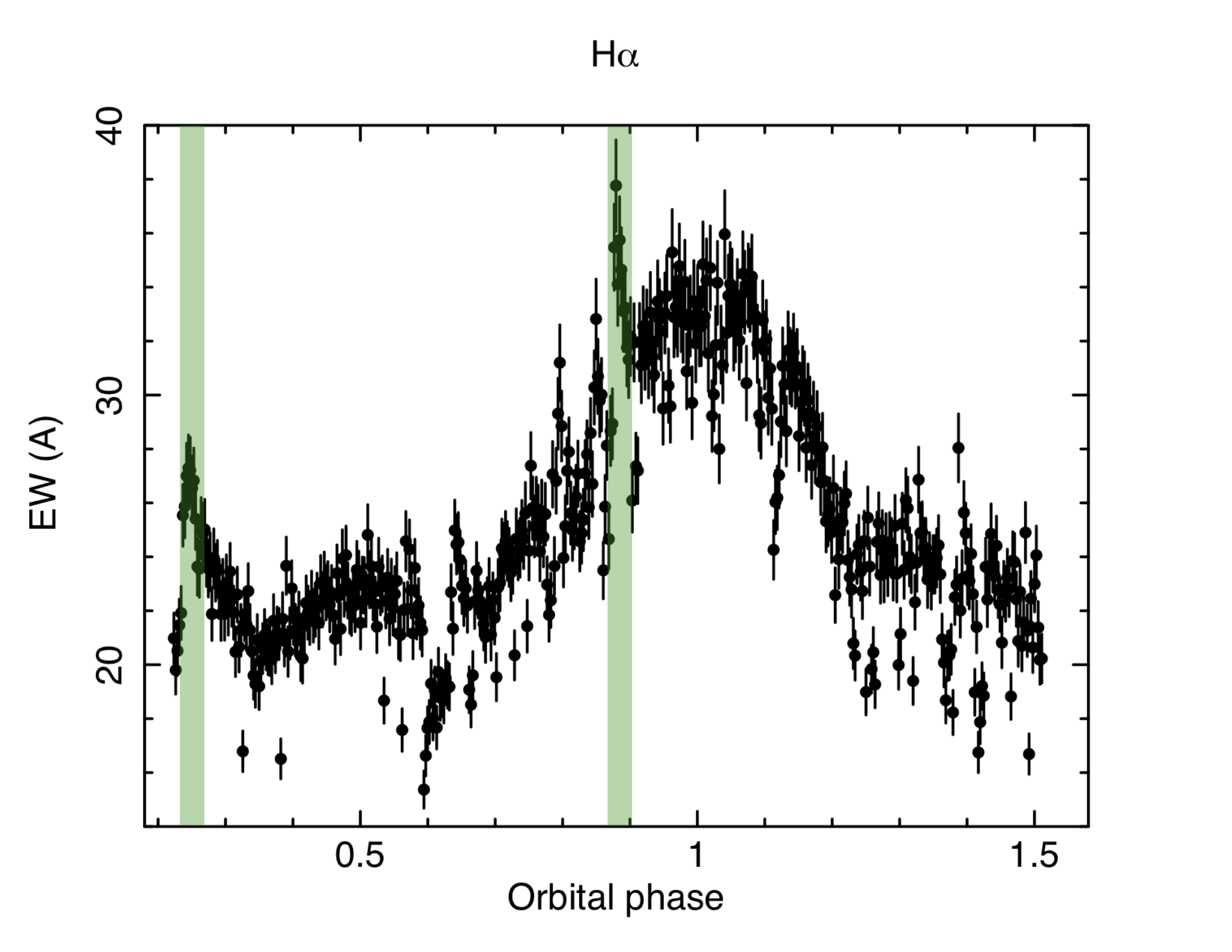}} \quad
{\includegraphics[width=.315\textwidth]{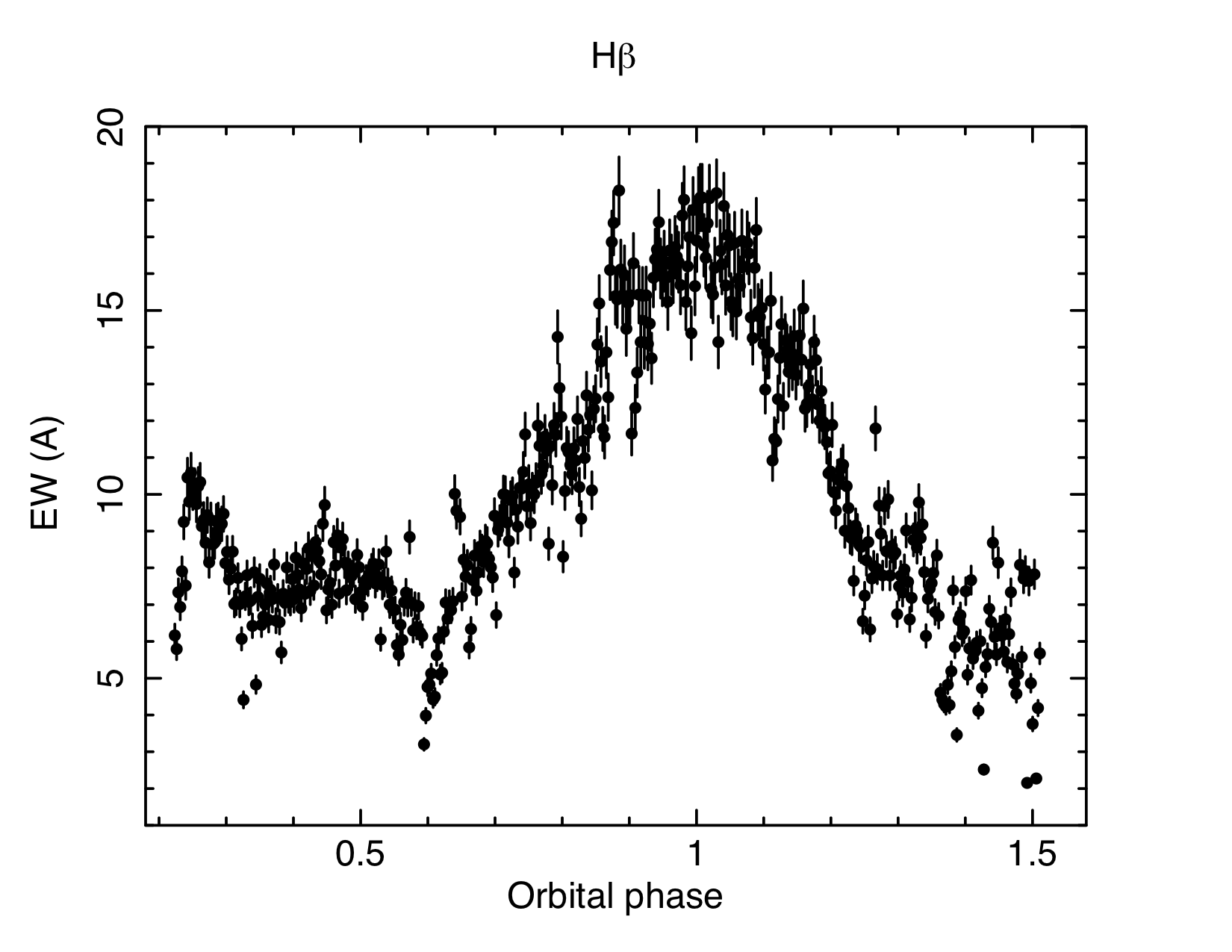}}\quad
{\includegraphics[width=.315\textwidth]{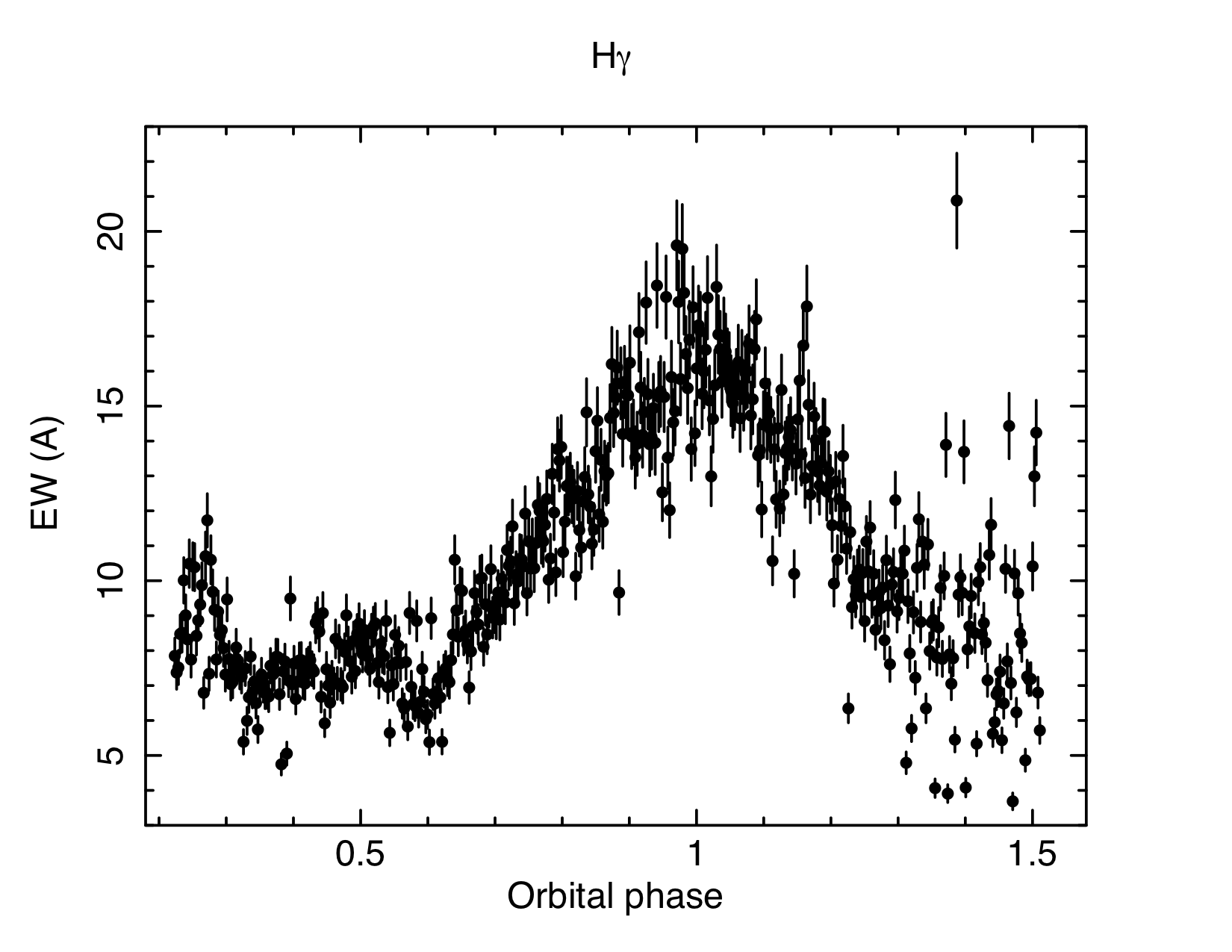}} \\

{\includegraphics[width=.32\textwidth]{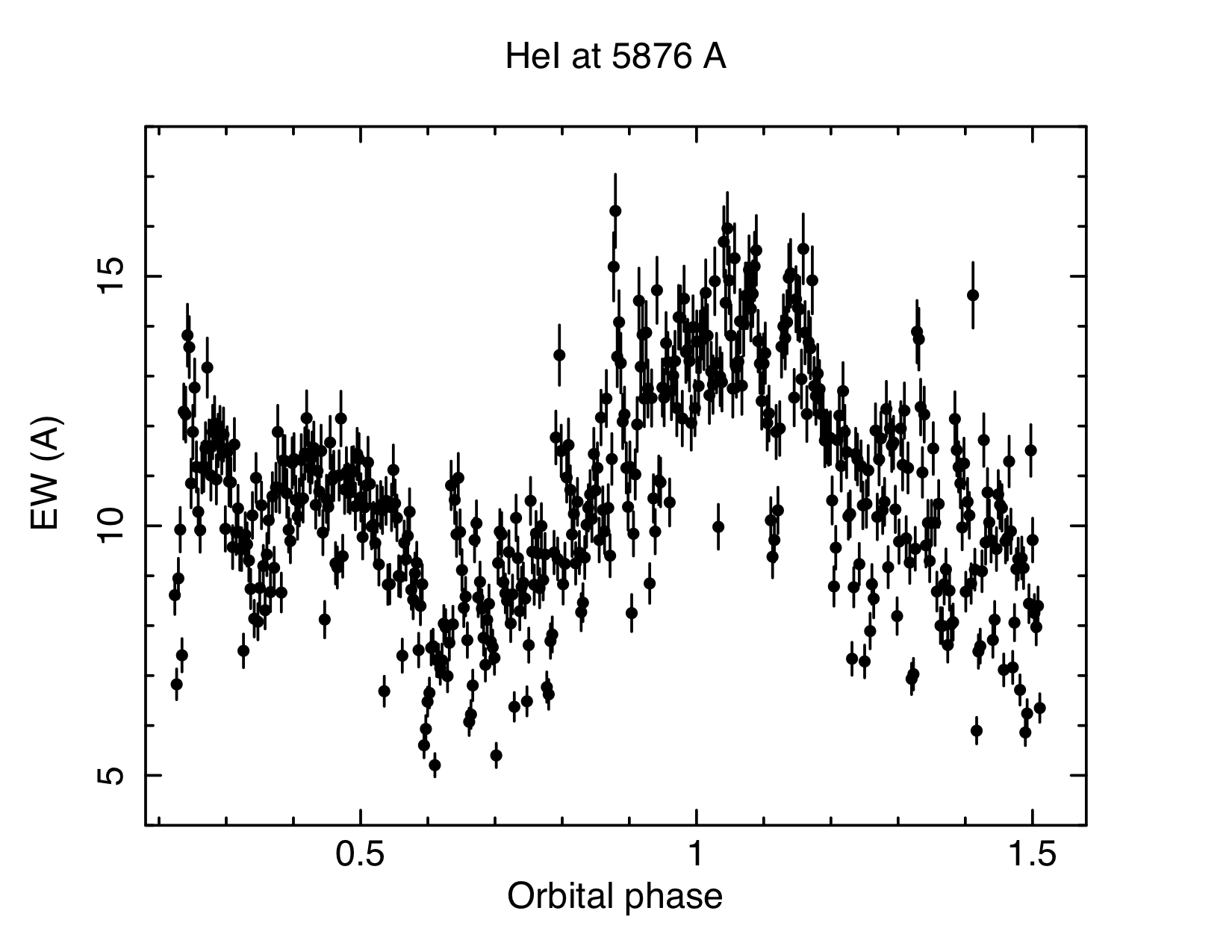}}\quad
{\includegraphics[width=.32\textwidth]{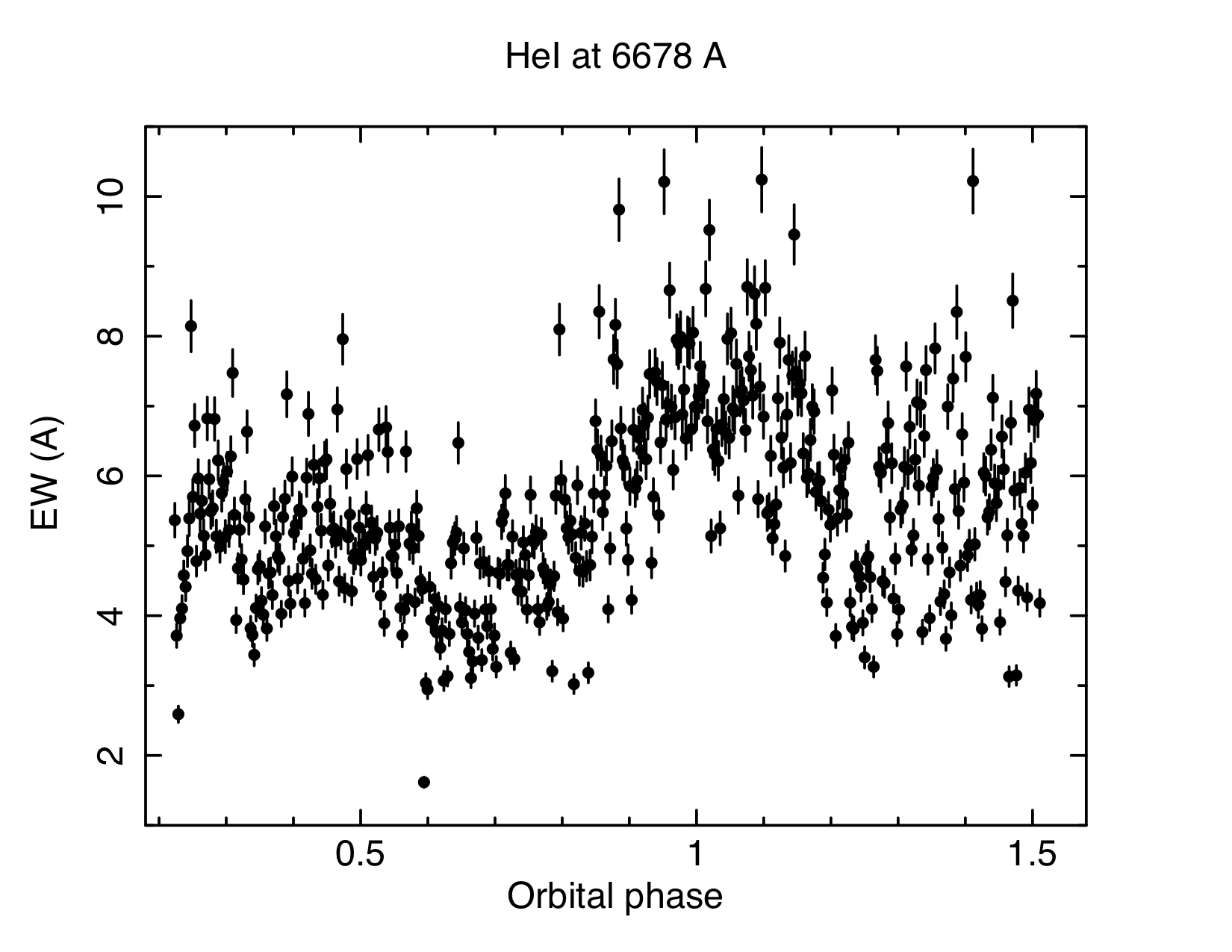}}\quad
{\includegraphics[width=.32\textwidth]{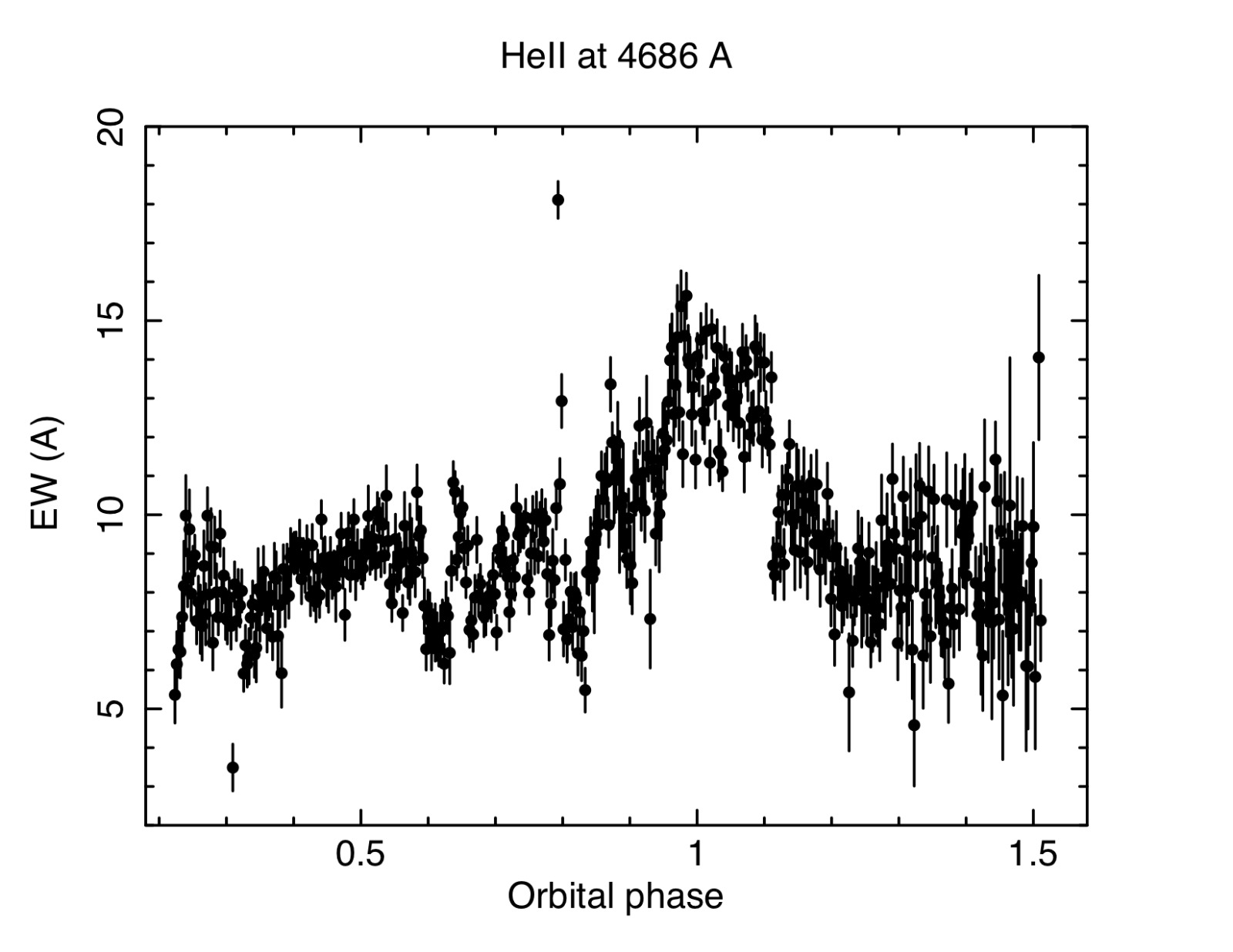}}

\caption{Equivalent width as a function of orbital phase for the H$\alpha$, H$\beta$, H$\gamma$, HeI at 5876 $\AA$, HeI at 6678 $\AA$ and HeII at 4686 $\AA$ emission lines. In green are underlined the two isolated episodes of enhanced emission described in Sec. \ref{Sec:EW}}
\label{fig:EWtutte}
\end{figure*}

To investigate the strength of the individual emission lines, we normalised the optical continuum and subsequently measured the equivalent width (EW)\footnote{For the study presented here, the EW measurement was made through the \texttt{light} task under the \texttt{molly} package.} of the strongest features. Fig. \ref{fig:EWtutte} shows the EW evolution as a function of the orbital phase for the most prominent emission lines. Note that, in this case, the seeing will not affect the line characteristics. As illustrated in the first panel, which displays the behaviour of the H$\alpha$ emission line, the EW exhibits a pronounced maximum centred at orbital phase $\simeq$ 1. In addition, the distribution suggests the presence of a local maximum at the beginning of the observation and around phase $\simeq$ 0.5, together with two drops in line strength at phases $\simeq$ 0.6 and $\simeq$ 1.1. Interestingly, two isolated episodes of enhanced emission appear to be present around phase $\simeq$ 0.2 and phase $\simeq$ 0.9 (highlighted in green in the figure). A similar behaviour is observed in the other emission lines, although less pronounced owing to their lower line intensities. Given that the line EW is less affected by seeing variations, we did not apply any seeing filtering in this analysis.

\noindent

\subsection{Full width at half maximum}
\begin{figure*}[h!]
\centering
    {\includegraphics[width=0.31\textwidth]{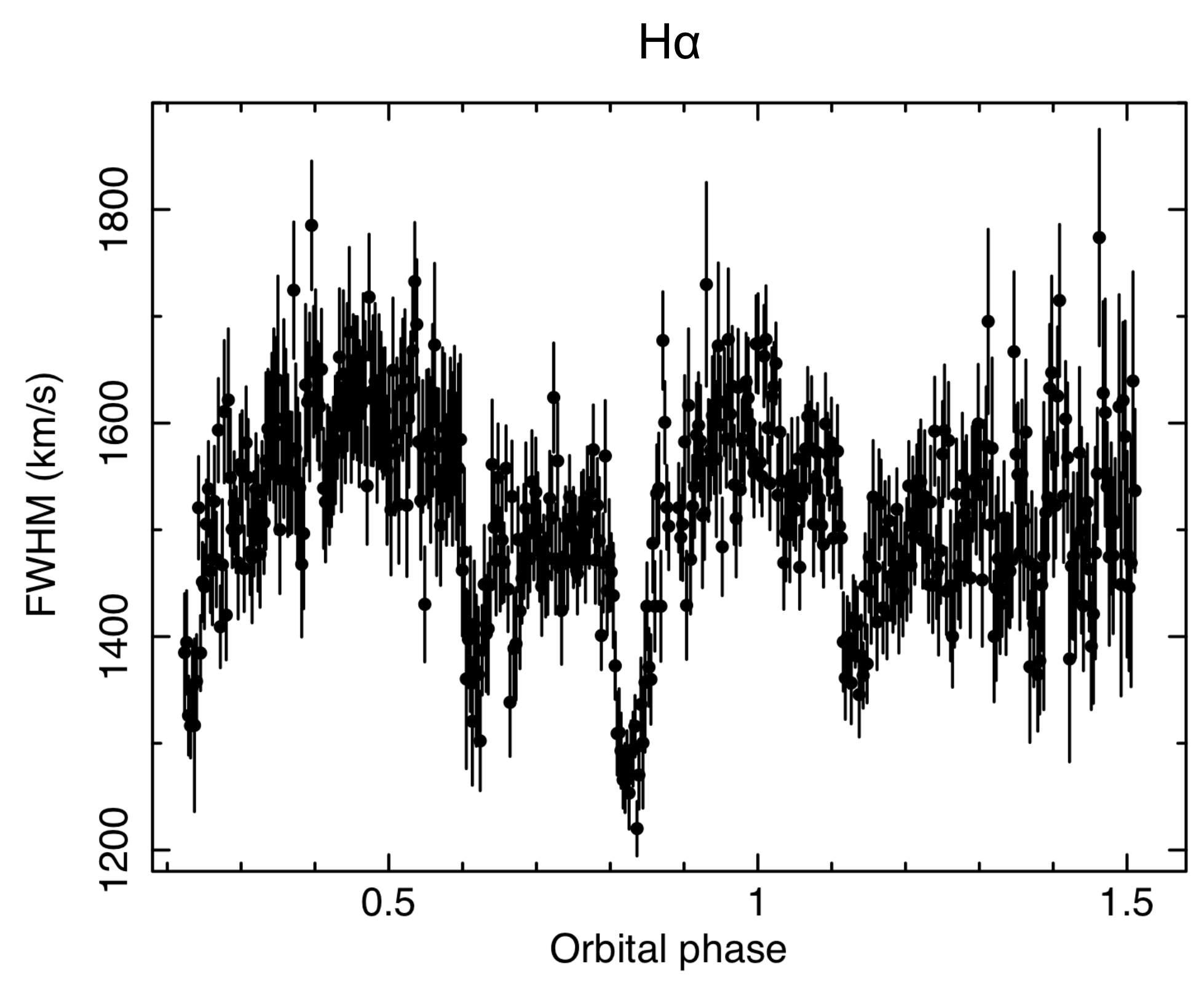}} \quad
    {\includegraphics[width=0.31\textwidth]{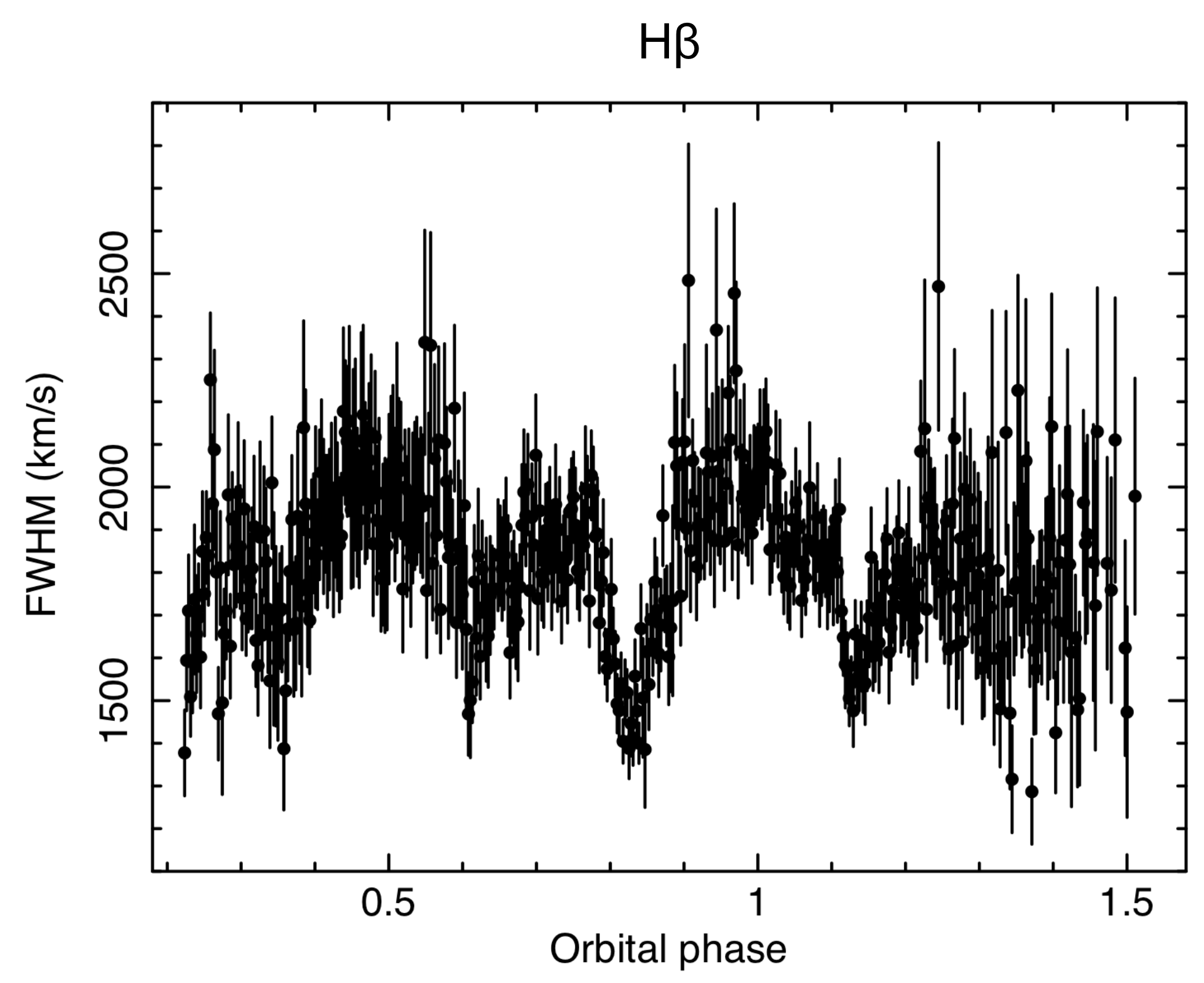}} \quad
    {\includegraphics[width=0.31\textwidth]{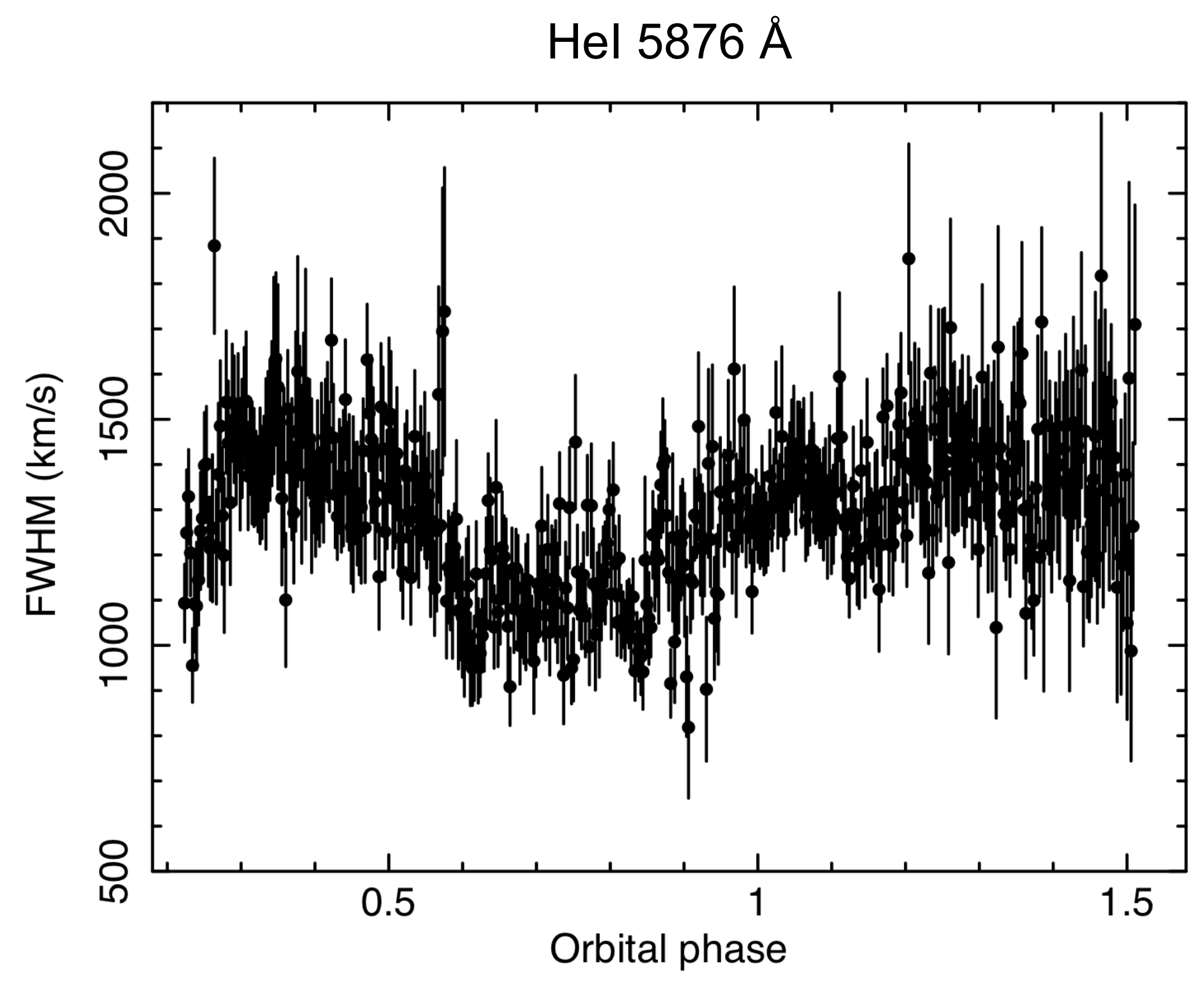}}
\caption{Three plots showing, respectively, the trend of the FWHM for H$\alpha$, H$\beta$, and HeI at 5876 $\AA$ emission lines over the orbital phase.}
\label{fig:FWHMtutte}
\end{figure*}

An additional parameter used to investigate the characteristics of the emission lines is the full width at half maximum (FWHM), which provides information on the velocity structure of the accretion disc. To measure the FWHM, we fitted the wings of the emission lines masking the central core with a model consisting of a constant plus a broad Gaussian component. This procedure was applied to the strongest lines, namely H$\alpha$, H$\beta$, and HeI at 5876 $\AA$. Figure \ref{fig:FWHMtutte} shows the FWHM evolution as a function of orbital phase for these three lines. As can be seen from the first panel, illustrating the H$\alpha$ line, the FWHM displays significant variability, with maxima around orbital phase $\simeq$ 0.4 and $\simeq$ 1 but also minima at the beginning of the observation, around phases $\simeq$ 0.6 and $\simeq$ 0.8, and at phase $\simeq$ 1.1. Interestingly, the minimum at phase $\simeq$ 0.6 appears to coincide with the decrease observed in the EW, and some of the FWHM variations correlate with features seen in the EW evolution; a more detailed discussion of this behaviour is presented in Section \ref{Par:discussion}.

\noindent

\subsection{Doppler Tomography}
\begin{figure*}[h!]
\centering
{\includegraphics[width=.31\textwidth]{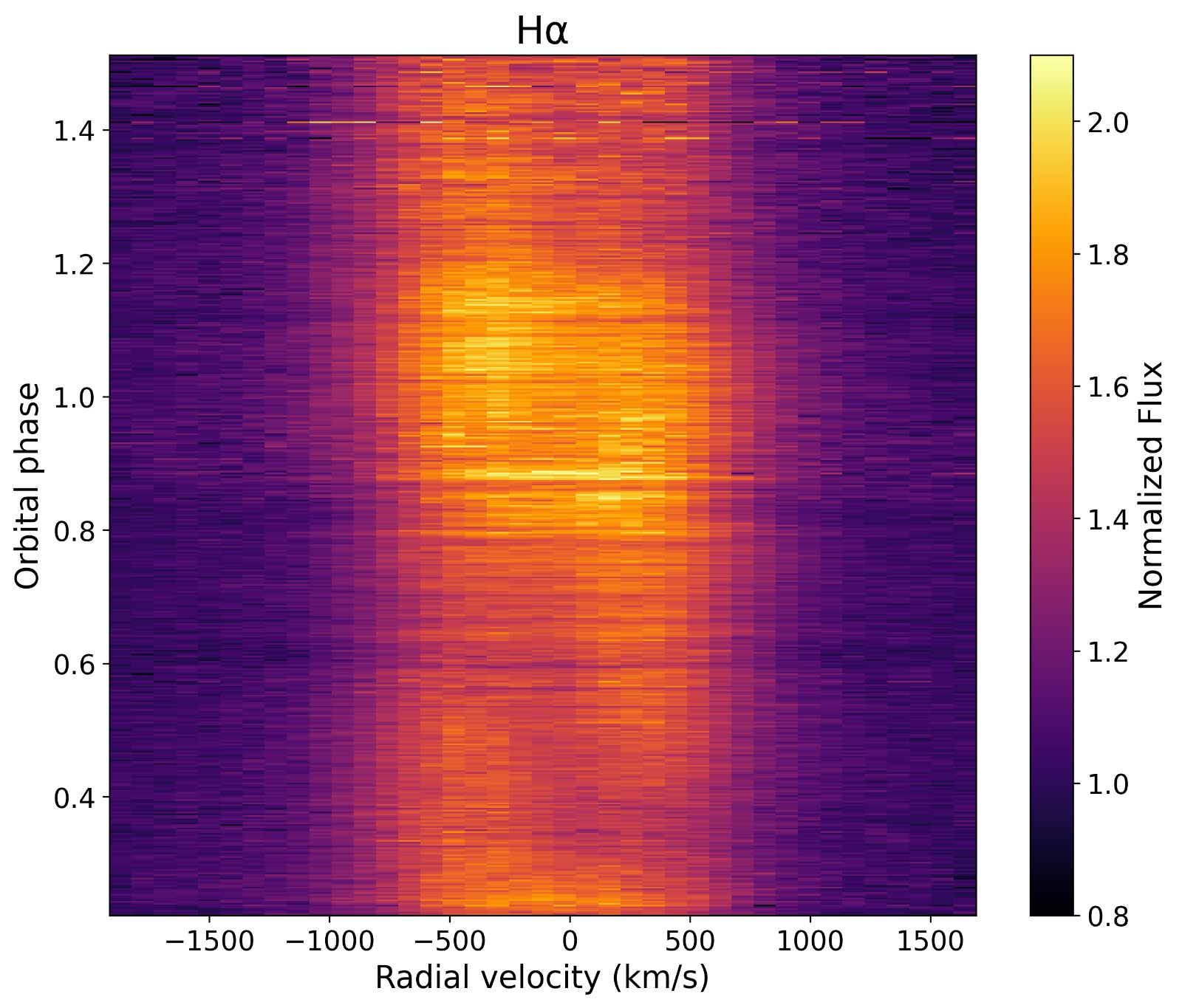}} \quad
{\includegraphics[width=.31\textwidth]{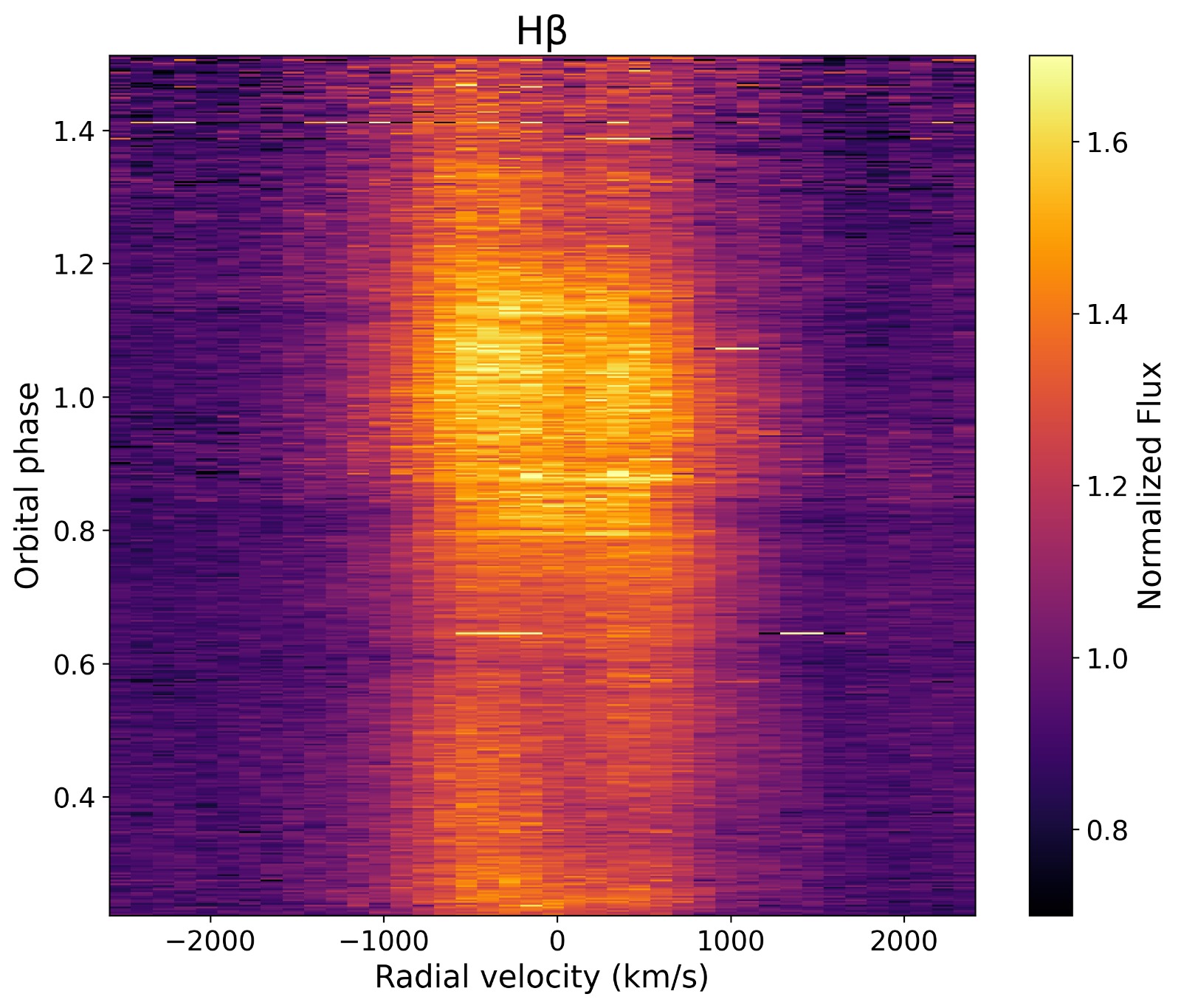}} \quad
{\includegraphics[width=.31\textwidth]{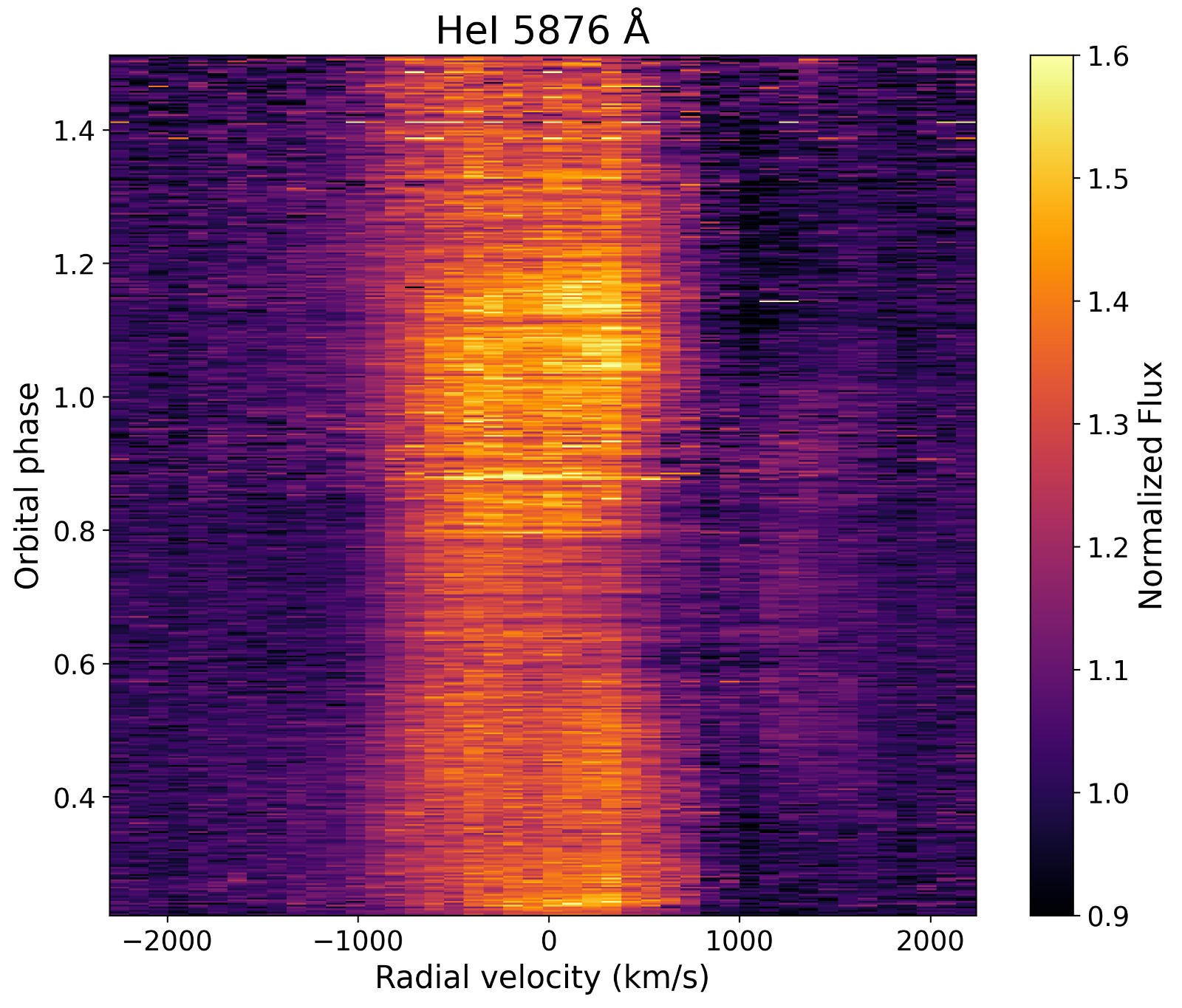}} 
\caption{Trailed spectrograms for the H$\alpha$, H$\beta$ and HeI at 5876 $\AA$ emission lines.}
\label{fig:Trail}
\end{figure*}

\begin{figure*}[h!]
\sidecaption
    \includegraphics[width=12cm]{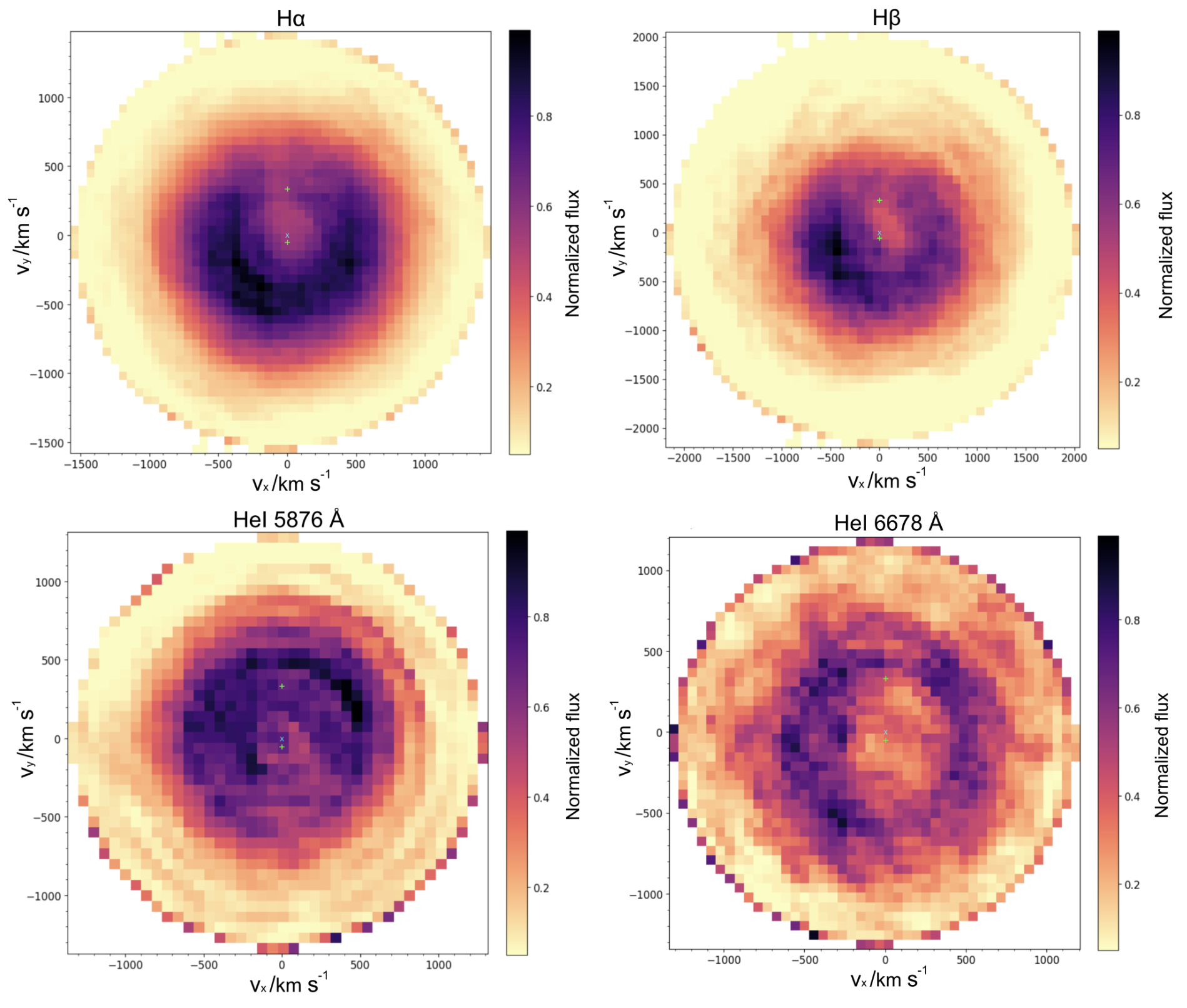}
        \caption{Doppler images in the velocity space for H$\alpha$ (top left), H$\beta$ (top right), HeI at 5876 $\AA$ (bottom left) and HeI at 6678 $\AA$ (bottom right) emission lines. The crosses indicate the positions of the compact object and the companion star, while the X marks the origin of the coordinate system.}
        \label{fig:Doppler}
\end{figure*}

We constructed a trailed spectrogram for the main emission lines to inspect their variability and compare it with what we observed for the EW (Fig. \ref{fig:Trail}).
\noindent
With the aim of exploring the variability observed in the EW profile, we performed Doppler tomography (\citealt{Marsh1988}; \citealt{Marsh2001}) on the main emission lines, trying to construct the accretion structures in the binary system. This technique reconstructs an effective map of the emission brightness in velocity space from phase-resolved spectra, allowing different emission components, such as those originating from the accretion disc or the secondary star, to be disentangled in velocity space. Through the \texttt{pydoppler}\footnote{\url{https://github.com/Alymantara/pydoppler}} package, it was possible to upload the spectra, select the range of wavelength for the emission line and for the background continuum and then generate the map, assuming an inclination i $\sim$ 54$^{\circ}$, a mass ratio of 0.137, a semi amplitude of the primary of 38 km \ s$^{-1}$, a systemic velocity of -90 km\ s$^{-1}$ and a mass of the primary of 1.7 M$_{\odot}$ \citep{McConnell2015,Shahbaz2019}. In Fig. \ref{fig:Doppler} we show the Doppler maps for H$\alpha$, H$\beta$, HeI at 5876 $\AA$ and HeI at 6678 $\AA$ emission lines.

\section{Discussion}\label{Par:discussion}
The previous sections have shown that the properties of the emission lines of J1023 in its sub-luminous state exhibit significant variations among all the orbital phases. All X-ray observations of J1023 in literature have shown that, during the sub-luminous accretion disc state, the source exhibits pronounced short-timescale variability, characterised by switches between the well-defined low and high modes. These transitions give rise to a clear bimodal distribution of the observed count rates, which is particularly evident in the X-ray light curve, but also observed in optical/NIR \citep[e.g.,][]{Shahbaz2015,Linares2014,Bogdanov2015,CotiZelati2018,BaglioZelati2023}. 
In the absence of simultaneous X-ray observations that would allow us to directly identify the times of these mode switches, it is therefore interesting to investigate whether signatures of such variability can be traced in the optical properties of the source, such as the emission-line behaviours.

\subsection{Continuum variability}
The observed variability of the optical continuum (Fig. \ref{fig:continuum-fit}) appears to indicate the presence of an orbital modulation in the light curve likely associated with emission from the irradiated companion star and the accretion disc. This is supported by the fit with a model consisting of a constant plus a sinusoidal component, which provides a statistically significant improvement ($>8\, \sigma$) over a constant-only model, as indicated by the F-test. The best-fit sinusoid is consistent with a maximum at orbital phase $\simeq$ 0.5, corresponding to the best visibility of the accretion disc and irradiated face of the companion.
Previous studies of J1023 have reported both the presence \citep{CotiZelati2014} and the absence \citep[or only a weak detection]{Messa2024} of the photometric orbital modulation. In this context, the relatively poor atmospheric conditions during our observations likely limit our ability to fully characterise the modulation. Nevertheless, the observed modulation amplitude appears to be approximately a factor of two lower than that reported by \cite{CotiZelati2014}. Inspection of the fit residuals does not reveal clear evidence for the presence of distinct high and low modes. However, given the limitations inherent to slit spectroscopy photometry and the variable observing conditions, we cannot exclude that part of the residual variability may arise from mode transitions. We therefore limit our interpretation to the conclusion that the observed modulation is consistent with irradiation of the companion star, without attempting a quantitative determination of the heating efficiency or companion star temperature. Such analyses have been previously performed using dedicated photometric datasets, obtaining a companion star temperature of $T_\ast = 6128 \pm 33$ K (\citealt{Shahbaz2022}). Further observations obtained under better seeing conditions would therefore be required to better constrain the properties and stability of the modulation.
The non-detection or weak detection of this orbital modulation in previous works has been interpreted as evidence for an increased contribution from the accretion disc in the optical band \citep[e.g.][]{BaglioZelati2023}. As the disc refills, its emission may become dominant diluting the irradiated companion. This interpretation was also considered for the observations of the accreting millisecond pulsar XTE J1814–338, which exhibited a similar behaviour during its quiescent state \citep{Baglio2013}.
When examining the possible correlations among the continuum, EW, and FWHM, no clear overall common behaviour is found. This result is supported by the Pearson and Spearman correlation coefficients computed for these quantities for the two strongest lines, H$\alpha$ and H$\beta$ (Table \ref{tab:correlazioni}), with the exception of the trend between the continuum and the EW. To further investigate possible correlations, we examined the variability of the different spectral-line properties individually.\\
Considering only the blue points in the colour–magnitude diagram (i.e. those obtained under seeing conditions consistent with the slit width, Fig. \ref{fig:color-magnitude}), a clear trend emerges in which the colour increases towards fainter magnitudes, corresponding to larger Be-Re values. This indicates that the spectra become progressively redder as the source decreases in brightness. The same behaviour is further reinforced when including spectra obtained under slightly worse seeing conditions (between 1.5 and 2.1 arcsec; orange points). A similar behaviour has been reported for candidate transitional millisecond pulsars such as 3FGL J1544.6-1125 and CXOU J110926.4-650224 (\citealt{Coti2024}; \citealt{Kennedy2020}; \citealt{Illiano2025}). In those systems, the reddening has been interpreted as a consequence of changes in the inner accretion flow, such as a partial evacuation of the inner disc. Therefore, while a similar physical mechanism may be at play, the observed trend could more generally reflect changes in the relative contribution of different emission components (e.g. outer disc, inner flow).

\begin{table}[h!]
\caption{\label{t7}Pearson and Spearman test values for the correlations.}
\centering
\begin{tabular}{ccccc}
\toprule
\begin{tabular}[c]{@{}c@{}}\textbf{Pearson}\\ {}\end{tabular} & 
\begin{tabular}[c]{@{}c@{}}\textbf{P-value}\\ {}\end{tabular} &
\begin{tabular}[c]{@{}c@{}}\textbf{Spearman}\\ {}\end{tabular} &
\begin{tabular}[c]{@{}c@{}}\textbf{P-value}\\ {}\end{tabular} &
\begin{tabular}[c]{@{}c@{}}\textbf{Line}\\ {}\end{tabular}\\
\midrule
\multicolumn{5}{c}{Be band over EW value}\\
\midrule
-0.772 &0.000  & 0.765 &0.000      &H{\fontsize{3mm}{4mm}\selectfont $\alpha$}\\
-0.748  &0.000 & -0.740 &0.000       &H{\fontsize{3mm}{4mm}\selectfont $\beta$}\\
\midrule
\multicolumn{5}{c}{Be band over FWHM value}\\
\midrule
0.194 &0.001  & 0.174 &0.005      &H{\fontsize{3mm}{4mm}\selectfont $\alpha$}\\
0.154  &0.013 & 0.129 &0.037       &H{\fontsize{3mm}{4mm}\selectfont $\beta$}\\
\midrule
\multicolumn{5}{c}{FWHM over EW value}\\
\midrule
0.068 &0.137  &0.012 &0.792     &H{\fontsize{3mm}{4mm}\selectfont $\alpha$}\\
0.160 &0.001 & 0.132 &0.005     &H{\fontsize{3mm}{4mm}\selectfont $\beta$}\\
\bottomrule

    \end{tabular}
    \tablefoot{The top panel shows the Pearson and Spearman results with the respective P-value for correlations between the continuum flux in the Be band (obtained for the spectra with limited seeing $<$ 2.1 arcsec), EW and FWHM of the H$\alpha$ and H$\beta$ emission lines.}
    \label{tab:correlazioni}
\end{table}

\noindent

\subsection{FWHM comparison}
\begin{figure}[h!]
   \centering
   \includegraphics[width=1\linewidth]{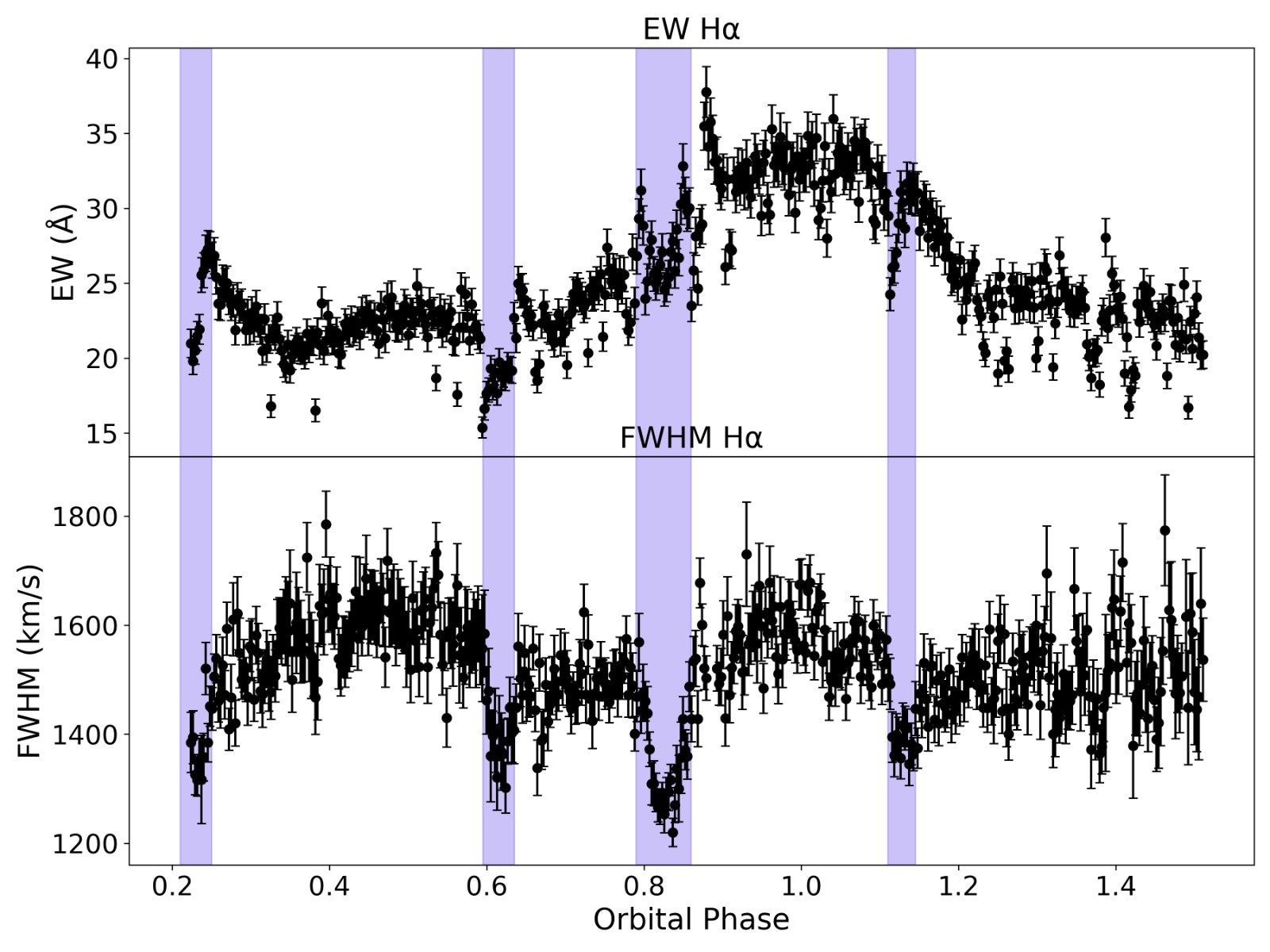}
      \caption{Equivalent width (EW; top panel) and full width at half maximum (FWHM; bottom panel) of the H$\alpha$ line as a function of orbital phase. The two quantities are shown on the same phase scale for direct comparison. The blue shaded regions highlight the phase intervals where the FWHM profile exhibits local minima.
              }
         \label{fig:EW_FWHM_comparison}
   \end{figure}

The FWHM (which reflects the velocity distribution of the emission regions) of the H$\alpha$ line, for instance (Fig. \ref{fig:FWHMtutte}, left panel), exhibits four well-defined minima, occurring at the beginning of the observation $\simeq$ 0.22, around orbital phases $\simeq$ 0.60, $\simeq$ 0.85, and $\simeq$ 1.15. 
The first two FWHM minima seem to be associated with a decrease of the EW of H$\alpha$, with a similar behaviour observed for weaker lines such as H$\beta$, H$\gamma$, and the He I transitions (Fig. \ref{fig:EWtutte}, for simplicity, we focus here on H$\alpha$). For the third minimum the EW shows only a gradual decrease without displaying a well-defined minimum. In contrast, at the orbital phase corresponding to the last FWHM minimum, the EW exhibits a short-lived minimum followed by a rapid change in slope, with values increasing thereafter (as can be seen in the Fig. \ref{fig:EW_FWHM_comparison}).
\noindent
This comparison suggests that the observed FWHM minima may originate from two different mechanisms. In cases where a correspondence with the EW is present, a reduced contribution of the region at higher velocity appears to be accompanied by a reduction in the emission line strength, possibly indicating that the contribution from the higher velocity emitting material, presumably located in the inner regions of the disc, is temporarily suppressed or reduced at those orbital phases. If these phases-epochs correspond to low modes, this behaviour would be consistent with a scenario in which transitions from high to low modes are associated with the ejection of the innermost disc regions (assuming that optical low modes correspond to X-ray low modes). Conversely, for the FWHM minima lacking an EW counterpart, a plausible interpretation is that emission from the cooler, lower-velocity regions of the disc dominates over that from the inner disc.

\subsection{EW profile}
\begin{figure}[h!]
   \centering
   \includegraphics[width=1\linewidth]{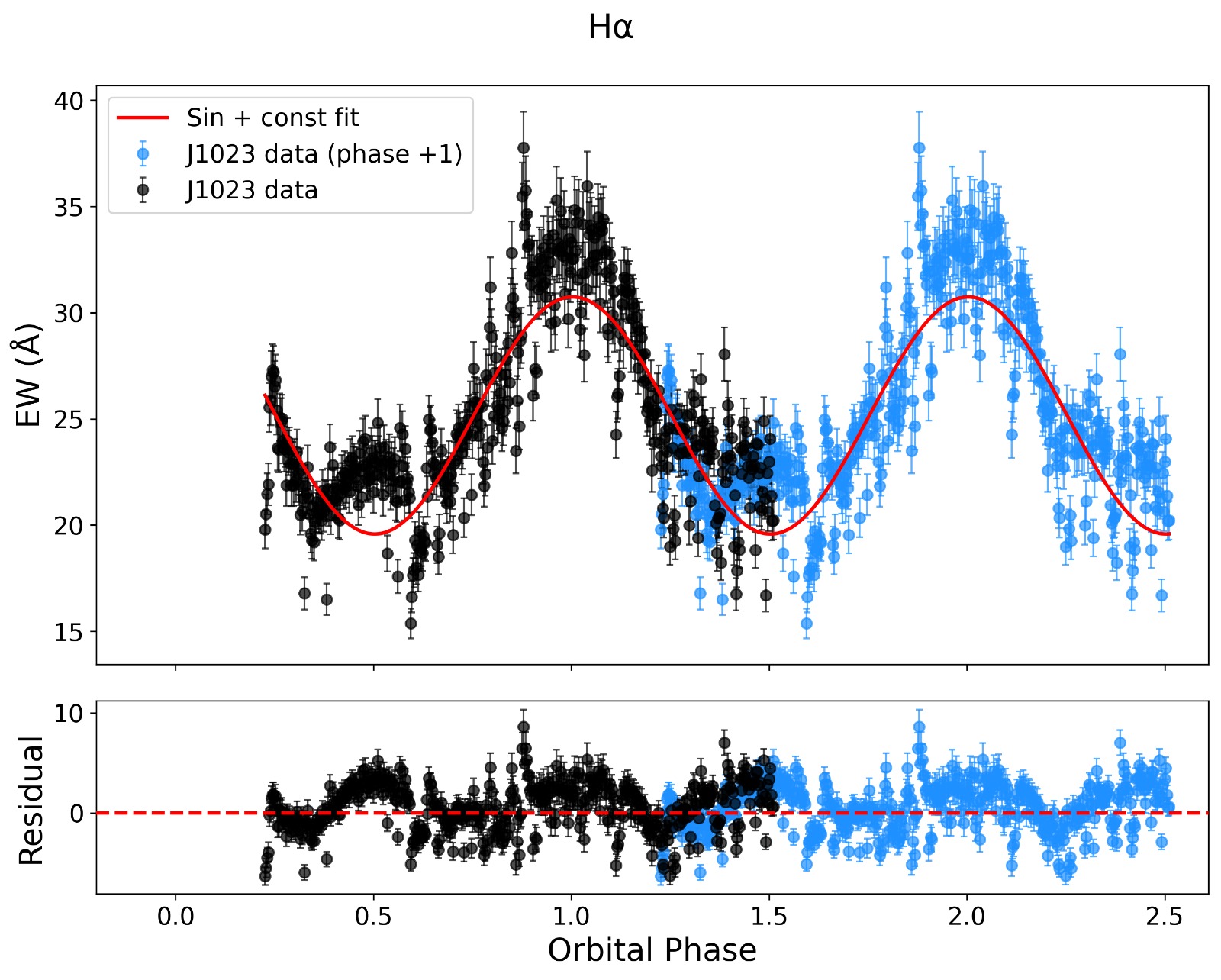}
      \caption{Equivalent width of H$\alpha$ as a function of orbital phase. The black points show the observed data, while the blue points represent the same measurements shifted by +1 in orbital phase to highlight possible periodic patterns. The red curve shows the best-fit sinusoidal model with a fixed period at the orbital period. The bottom panel displays the residuals of the fit.
              }
         \label{fig:Residual}
   \end{figure}

An inspection of the EW evolution (Fig. \ref{fig:EWtutte}) reveals, in addition to a broad and pronounced maximum centred at orbital phase $\simeq$ 1, the presence of a possible narrow flare-like event around phase $\simeq$ 0.9. This feature is short-lived, spanning approximately seven spectra ($\sim$ 7 minutes), and does not appear to be associated with any distinctive behaviour in the FWHM (Fig. \ref{fig:EW_FWHM_comparison}). 
With a pronounced EW maximum observed at orbital phase $\simeq$ 1 (the phase at which the neutron star and hence the accretion disc is at superior conjunction, Fig. \ref{fig:EWtutte}), the EW modulation is strongly anti-correlated with the optical continuum modulation (Fig. \ref{fig:continuum-fit}). At this phase, the non-irradiated face of the companion star is oriented towards the observer. Since the optical light curve is dominated by the irradiated face of the companion at phase $\simeq 0.5$, continuum dilution is the most natural explanation for the EW minimum at that phase. Nevertheless, changes in EW reflect a combination of continuum variations, line intensity, and FWHM, and contributions from these effects, as well as from the phased continuum of the accretion disc, cannot be entirely excluded. The observed anti-correlation between the sinusoidal light curve and the EW profile (Tab.~\ref{tab:correlazioni}) is consistent with this picture. Concerning the H$\alpha$ line flux, some variability is indeed present over the orbital phases at a level of a few percent.  
To further investigate the observed behaviour, we fitted the EW evolution with a model consisting of a constant plus a sinusoidal component (Fig. \ref{fig:Residual}). In a first fit, the period was fixed to the orbital value (P = 4.75 hr), while in a second fit, it was left free. In both cases, the resulting parameters are consistent with a sinusoidal modulation, with an average EW of $\simeq$ 24.8 $\AA$ and a semi-amplitude of $\simeq$ 4.9 $\AA$, and a modulation compatible with the orbital period. The fit therefore suggests that the observed sinusoidal trend, with a maximum around orbital phase $\simeq$ 1, may reflect a periodic behaviour. In particular, the EW semi-amplitude ($\simeq$20\% of the mean value) is comparable to that observed in the Re light curve, suggesting that continuum dilution associated with the irradiated companion star plays an important role in shaping the observed EW behaviour. In this framework, the EW curve resembles an inverted heating modulation, with a minimum around orbital phase $\simeq$0.5 and a maximum around phase $\simeq$1.\\
To investigate this possibility, we compared the H$\alpha$ EW values with those obtained during the 2021 observing campaign presented in \cite{Messa2024}, adopting the same orbital ephemeris to compute the orbital phases. This comparison does not provide evidence for a persistent periodicity in the EW behaviour. However, the observed variations may also be affected by transitions between high and low modes, whose occurrence and relative fraction are known to vary between different epochs. Additional observations covering at least two full orbital cycles would therefore be required to better assess the presence of a recurrent modulation.

\begin{figure}[h!]
   \centering
   \includegraphics[width=1\linewidth]{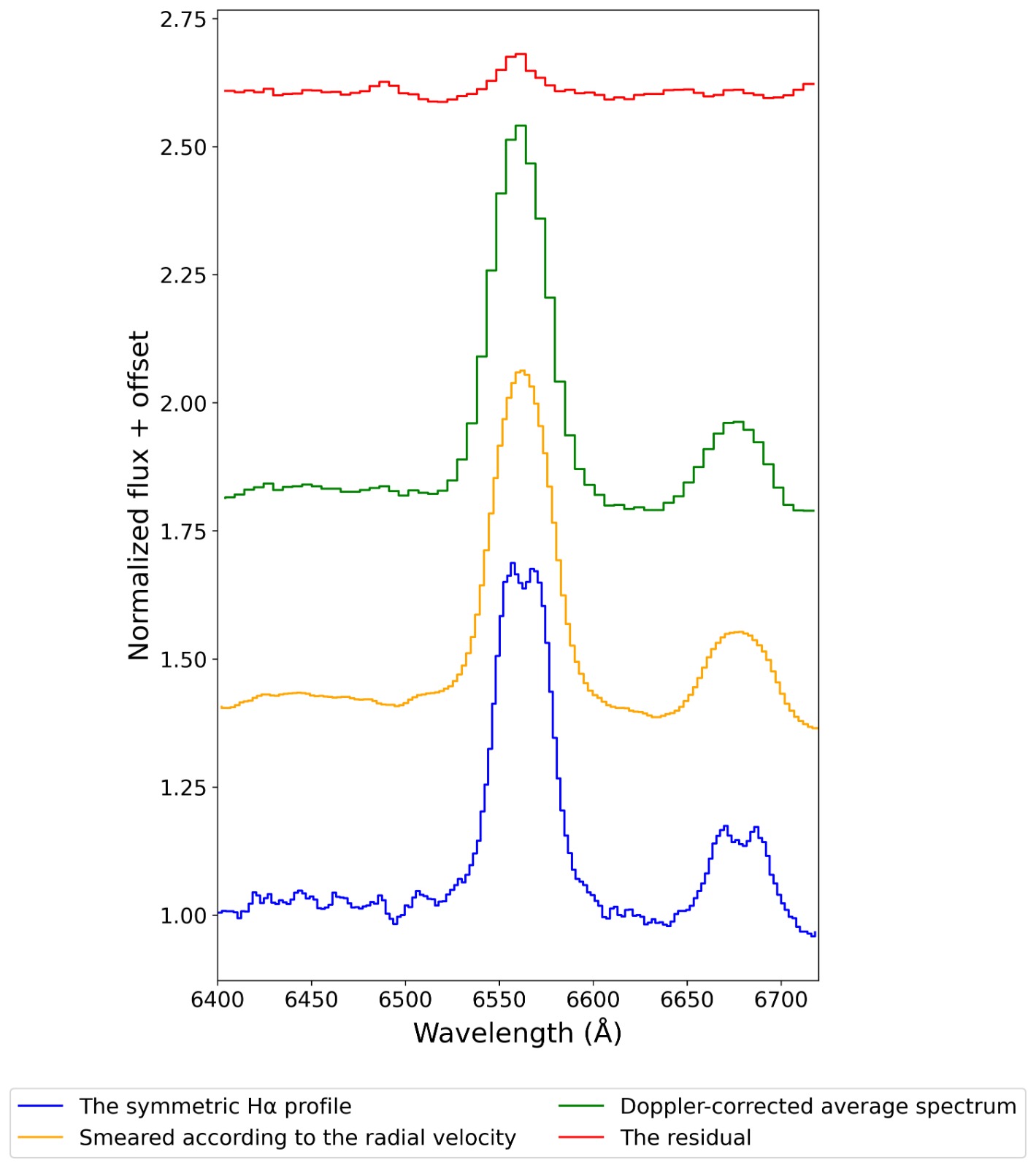}
      \caption{Removal of the narrow H$\alpha$ component. The symmetric H$\alpha$ disc profile simulated from the data (blue) is smeared according to the orbital phases of the individual spectra (yellow). Subtracting this profile from the Doppler-corrected average spectrum of J1023 in the rest frame of the secondary star (green) isolates the residual narrow H$\alpha$ component (red). Arbitrary vertical offsets have been applied for clarity.}
         \label{fig:Torres}
   \end{figure}

\noindent
To assess a possible contribution from the companion star to the observed EW behaviour, we extracted the corresponding emission H$\alpha$ component from the emission line profiles following the method described by \cite{Casares1997} and \cite{Torres2002}. The procedure is illustrated in Fig. \ref{fig:Torres}.

\noindent
First, we modelled the contribution of the accretion disc by averaging all spectra covering exactly one orbital period in the laboratory rest frame. This approach allows one to obtain a symmetric double-peaked profile (blue spectrum in Fig. \ref{fig:Torres}), as using all available spectra would introduce artificial asymmetries, either blue or red shifted peaks, caused by the S-wave emission (the periodic variation in the position of the spectral line, due to the hot-spot at the rim of the accretion disc) when the phase coverage exceeds one orbital cycle. The resulting profile represents a stationary disc component without the S-wave contribution.
This symmetric disc profile was then replicated to match the number of observed spectra covering one orbital period and shifted into the reference frame of the companion star by correcting for its orbital motion. By averaging these shifted profiles, we obtained the mean disc spectrum in the companion frame (yellow spectrum in Fig. \ref{fig:Torres}).
The original spectra covering one orbital period were then corrected for the orbital motion of the companion and averaged, producing the spectrum shown in green in Fig. \ref{fig:Torres}. Subtracting the disc model (yellow spectrum) from this averaged spectrum results in the residual H$\alpha$ component (red spectrum), which has a FWHM value of 1050.6 $\pm$ 292.4 km \ s$^{-1}$ and a radial velocity of -34.9 $\pm$ 14.6 km \ s$^{-1}$. This procedure was carried out using the \texttt{molly} package. The resulting spectrum appears to indicate a modest contribution from emission originating from the companion star given that, under such scenario, we would have expected a much narrower line profile. \\
To further assess this result and to investigate the EW behaviour in more detail, it is useful to compare these findings with the Doppler tomography maps shown in Fig. \ref{fig:Doppler}. 
If a significant contribution from emission originating on the companion star were present, one would expect a concentration of flux at velocities close to v$_x$ $\simeq$ 0 and positive v$_y$ (in this case around 300 km \ s$^{-1}$) in Doppler velocity space (\citealt{Steeghs1998}, Fig. 1; \citealt{Torres2002}, Fig. 3). No such feature is observed in any of the four Doppler maps.
In the case of the H$\alpha$ and H$\beta$ map in particular, emission appears to be largely absent from this region. These results therefore indicate that the pronounced EW maximum observed at orbital phase $\simeq$ 1 is unlikely to originate from emission associated with the companion star, and its origin requires further investigation.

\subsection{Doppler maps}
The Doppler maps (Fig.~\ref{fig:Doppler}) generally display the expected torus-like morphology associated with emission from an accretion disc. However, a more careful visual inspection of the Balmer Doppler maps reveals that the dominant emission is distributed over negative v$_x$ and v$_y$ velocities, extending into the quadrant of positive v$_x$ and negative v$_y$. The predominance of emissivity in these quadrants closely resembles the Doppler map reported for the neutron star low-mass X-ray binary XTE J2123–058 by \citet[see their Fig.~3]{Casares2002}. As a robustness test, we reconstructed the Doppler maps after excluding the spectra corresponding to the sharp FWHM variations highlighted in Fig. \ref{fig:EW_FWHM_comparison}. The resulting maps are essentially unchanged and display the same overall emission morphology as those obtained from the full dataset. This indicates that these short-lived events do not significantly affect the tomographic reconstruction, which therefore primarily reflects the average distribution of the emitting material over the orbital cycle. In \citealt{Casares2002}, the authors noted similarities with the behaviour observed in SW Sex-type cataclysmic variables, a class of nova-like interacting binaries characterised by high mass-transfer rates. These similarities suggest that a comparable accretion flow structure might be at work.\\
Indeed, part of the H$\alpha$ emission shown in Fig. 3 of \cite{Casares2002} arises at velocities that are too low to be readily associated with Keplerian motion in the accretion disc, as previously discussed by \cite{Hynes2000}. For SW Sex systems, as well as in the scenarios proposed by \cite{Hynes2000} and \cite{Casares2002}, this behaviour has been interpreted in terms of a propeller mechanism (\citealt{Wynn1997}). In this framework, a magnetic field anchored in the disc interacts with the accretion flow from the secondary star, expelling clumps of material from the system. While the trajectories of these clumps depend on their mass, they intersect in a restricted region of velocity space, where collisions between clumps of different sizes can give rise to the enhanced and asymmetric emission observed (\citealt{Hoard&Szkody1997}; \citealt{Horne1999}; \citealt{Hellier2000}).\\
In our case the Doppler maps were obtained during a low-accretion phase (i.e. the sub-luminous state), when the accretion rate is relatively low but rather similar to those of cataclysmic variables ($\dot{M}\sim$ 10$^{-12}$ - 10$^{-11}$ yr$^{-1}$, \citep{Archibald2015,Bogdanov2015,Papitto&Torres2015,Campana2016}). Under such conditions, the neutron star magnetosphere itself in principle may act as an efficient propeller (\citealt{Menou1999}), and strong outflows/winds in a dense accretion flow could further inhibit a large fraction of the material from reaching the neutron star (\citealt{Blandford&Begelman19999}). 
Doppler maps of the H$\alpha$ emission line for J1023 have recently been published. \cite{Hakala2018} analysed optical spectra obtained in February 2017, separating the data into flaring and non-flaring intervals based on the slit count rate. In the flaring state, they found that the emission is not fully axisymmetric, showing a deficit of emission at v$_y$ $\simeq$ 0 and positive velocities, as well as reduced emissivity in the quadrant with positive v$_x$ and negative v$_y$. They interpreted this lack of emission as originating from the region of the disc closest to the observer at orbital phases between $\sim$0.25 and $\sim$0.5. This feature was associated with material being expelled from the system through a propeller mechanism, consistent with the model presented by \cite{Wynn1997}.
Indeed, their Doppler map closely resembles the expected structure of a propeller-driven flow, showing contributions from the gas stream and emission concentrated in the quadrant with negative v$_x$ and v$_y$, invoking a re-impact of the expelled material onto the disc (see \citealt{Hellier2000} for a schematic view). 
Similarly, \cite{Shahbaz2019}, who analysed optical spectra obtained in March 2014 and February 2016, produced Doppler maps showing strong emission associated with the gas stream, indicative of an impact with the accretion disc (the hot spot). In their case, however, an excess of emission is observed at orbital phases between $\sim$0.25 and $\sim$0.5, a feature that is also present in both our H$\alpha$ and H$\beta$ maps (even if less intensely).
The differences between these previously published Doppler maps of J1023 and those presented in this work, most notably the absence of a clear gas stream emission component and the intensified emissivity in the quadrant with negative v$_x$ and v$_y$, may point to the same underlying propeller process operating under different disc conditions.\\
Although the Doppler maps may suggest the presence of a mechanism in which a fraction of the accreting material is expelled from the system, broadly consistent with a propeller-like scenario, several observational constraints in the case of J1023 challenge this interpretation. If the pulsar is active, the accreting material is expected to lie outside the light cylinder, while the magnetosphere remains within it. In contrast, a propeller regime would require matter to penetrate inside the light cylinder, a condition that in our case may only be met during low modes (\citealt{Veledina2019}), which occupy a relatively small fraction of the time ($\sim$20\%) and are therefore unlikely to dominate the observed Doppler maps. Moreover, a propeller scenario is typically associated with a high spin-down rate, which is not supported by observations. The spin-down rate measured by \cite{Burtovoi_2020} is consistent with that reported by \cite{Archibald2013} when the source was in the radio pulsar state. This suggests that the spin-down has remained essentially unchanged, making the presence of a dominant propeller regime unlikely. The interpretation of the Doppler maps therefore remains open, and a firm conclusion is hindered by both the observational constraints and the intrinsic complexity of the gas-stream patterns. Further investigation is therefore required to clarify which physical processes are at play and how they can give rise to the observed emission structures. In particular, observations spanning at least two orbital cycles and allowing for the identification of the different source modes would provide key constraints on the origin of these features.
Regarding the other Doppler maps shown in Fig. \ref{fig:Doppler}, the H$\beta$ line displays a pattern broadly similar to that observed for H$\alpha$, although with a lower overall intensity, particularly in the quadrants with positive v$_x$. In contrast, the two He I emission lines, at 5876 $\AA$ and 6678 $\AA$, exhibit markedly different structures. In the first case, an increase in emission is observed at positive v$_y$ velocities, with a possible contribution from the gas stream. In the latter, the Doppler map reveals a largely symmetric disc, with only localised regions of increased emissivity, because this line is produced at outer regions with respect to 5876 $\AA$. Overall, these maps closely resemble those presented by \cite{D'Avanzo2005} for the quiescent neutron star X-ray binary Centaurus X-4. Based on previous studies, one would generally expect higher velocities in the Doppler maps of helium lines compared to the Balmer lines, as He emission is typically associated with hotter, inner regions of a viscously heated accretion disc, where the temperature decreases with increasing radius. However, such a clear separation in velocity space is not observed in our case. This may indicate that the H$\alpha$ emission does not arise exclusively from the cooler, outer disc regions, but also has a significant contribution from higher-velocity material located closer to the inner disc. This interpretation is further supported by the trailed spectra and the Doppler maps of H$\gamma$, H$\delta$, and HeII at 4686 $\AA$ presented in Appendix \ref{appendix:B} (Fig. \ref{fig:Trail_appendice}, \ref{fig:Doppler_appendice}), which show similar trends. In particular, the Balmer Doppler maps display a more pronounced emission in the quadrant with negative v$_x$ and v$_y$, whereas the HeII map appears more symmetric, with a possible contribution from the gas stream.

\subsection{Stellar winds}
An alternative interpretation for the enhanced line emission observed at orbital phase $\simeq 1$ (Fig. \ref{fig:EW_FWHM_comparison}) may involve material associated with outflows or winds that, due to the orbital motion of the system, is directed toward the observer at specific orbital phases. In such a scenario, characteristic signatures would be expected in the emission-line profiles, such as P Cygni-type features (e.g. \citealt{Munoz2016}; \citealt{Munoz2022}; \citealt{Munoz2026}). We therefore examined the H$\alpha$ line profile in detail, searching for possible P Cygni absorption components or distinctive structures in the line wings. However, neither the inspection of individual spectra nor the analysis of spectra averaged over different orbital-phase intervals revealed clear signatures indicative of wind-driven emission. This is likely due to the spectral resolution of our data (6.5 \AA\ at H$\alpha$), which may not be sufficient to detect such features. Therefore, the present analysis does not rule out the possible presence of winds.

\section{Conclusions}\label{Sec::Conclusions}
We have presented optical high-time-resolution spectroscopic observations of the transitional millisecond pulsar PSR J1023+0038 obtained while the system was in the sub-luminous disc state. Our observations cover, for the first time, an entire orbital cycle of the binary system, providing a total of 480 optical spectra acquired with minute-timescale cadence. This represents the first optical spectroscopic study of a tMSP with such a high temporal resolution over a full orbital period. The data reveal significant variability in the properties of the main emission lines throughout the observation.\\
The main results of this work can be summarised as follows:

\begin{enumerate}
\item We find evidence for correlated behaviour between some of the minima observed in the FWHM and those detected in the EW of the main emission lines. This behaviour may suggest episodes of matter ejection from the innermost regions of the accretion disc, possibly associated with switches to low modes. However, simultaneous X-ray observations would be required to confirm the nature of these events.
\item The comparison of the H$\alpha$ and H$\beta$ Doppler maps presented in this work with those reported in the literature suggests the presence of asymmetric emission structures, possibly associated with the expulsion of part of the accreting material (as described in \citealt{Papitto2019} and \citealt{Veledina2019}). While this behaviour may be qualitatively consistent with a propeller-like scenario, several observational constraints in J1023 (e.g. the location of the accreting material with respect to the light cylinder and the relatively low spin-down rate) challenge this interpretation. In this context, an alternative mechanism capable of driving mass outflows could be associated with the pulsar wind.

\item We confirm the presence of variability in the main properties of the optical spectrum of J1023 on timescales of minutes. In addition, a longer-term variability, likely associated with the orbital period, is also observed. While the latter appears more structured, the short-timescale variability is likely of a more erratic nature. Further observational campaigns covering at least two orbital periods would help to better characterise the nature of these variations. 
\end{enumerate}
In addition, simultaneous multiwavelength observations would allow for a direct comparison between the variability of the main emission-line properties and the occurrence of high and low modes, thereby providing further insight into the physical origin of these episodes. Such observations would also enable the construction of Doppler maps for the different modes, offering a powerful diagnostic to probe the structure and dynamics of the accretion disc under varying accretion regimes.

\section{Data availability}
A copy of the reduced spectra are only available in electronic form at the CDS via anonymous ftp to \url{cdsarc.u-strasbg.fr} (130.79.128.5) or via \url{http://cdsweb.u-strasbg.fr/cgi-bin/qcat?J/A+A/}.

\begin{acknowledgements}
We thank the referee for helpful comments. This research is based on observations made with the Gran Telescopio Canarias (GTC), installed at the Spanish Observatorio del Roque de los Muchachos of the Instituto de Astrofísica de Canarias, on the island of La Palma (programme ID: GTC67-24A; PI: Francesco Coti Zelati). This work is partly based on data obtained with OSIRIS+, an instrument built by a consortium led by the Instituto de Astrofísica de Canarias in collaboration with the Instituto de Astronomía of the Universidad Autónoma de México. OSIRIS was funded by GRANTECAN and the National Plan of Astronomy and Astrophysics of the Spanish Government.
We thank the GTC staff for their support during the observing night, in particular Antonio Cabrera-Lavers, head of GTC operations. We also thank Alessandra Ambrifi for helpful discussions on the identification of signatures from pulsar winds.\\
M.C.B. acknowledges support from the INAF Fundamental Research Grant XBOOM.\\
F.C.Z. is supported by a Ramón y Cajal fellowship (grant agreement RYC2021-030888-I), by the Spanish grant ID2023-153099NA-I00, and by the program Unidad de Excelencia Maria de Maeztu CEX2020-001058-M.\\
DDM acknowledges support from INAF AF2022 FANS and AF2024 PULSE-X projects.\\
AR acknowledges financial support from the SOXS project (PI S. Campana) and from the PRIN-INAF 2022 "Shedding light on the nature of gap transients: from the observations to the models".\\
KA acknowledges financial support from the SOXS project (PI S. Campana).\\
\end{acknowledgements}

\bibliographystyle{aa}
\bibliography{bibliografia}

\appendix
\onecolumn
\section{Swift/XRT observations} \label{appendix:A}
The presence of emission lines in the optical spectra (Fig. \ref{fig:average}) is one of the main features to confirm tMSPs in the sub-luminous accretion state. This state is also characterised by peculiar multi-band properties, such as an X-ray emission that switches between high and low modes, a X-ray spectrum well described by an absorbed power-law with a photon index $\Gamma \sim 1.2-1.9$ and a typical luminosity of $\sim$$10^{33-34} \, \mathrm{erg \, s^{-1}}$. Due to the lack of simultaneous X-ray coverage during our optical campaign, we analysed archival Swift/XRT observations in photon counting (PC) mode performed before (March 12 and 13, 2024; Obs. IDs: 00033012226, 00033012227) and after our observations (April 10, 2024; Obs ID: 00033012233). This analysis aims to support the hypothesis that J1023 remained in this intermediate state throughout the period of our study. We analyzed the data using the online XRT product generator, based on HEASOFT v6.35.2.

The switches between high and low modes, together with sporadic flares, are evident in analysed X-ray light curves. Spectral fitting yielded neutral hydrogen column densities ($N_H$) of $10^{+8}_{-6}$ and $4^{+6}_{-4} \times 10^{20} \, \mathrm{cm^{-2}}$ for the March and April observations, respectively. During these two epochs, the X-ray spectrum was characterized by photon indices ($\Gamma$) of $1.7 \pm 0.2$ and $1.5 \pm 0.2$, with corresponding unabsorbed fluxes (0.3-10~keV) of $(1.0 \pm 0.1)$ and $1.3^{+0.2}_{-0.1} \times 10^{-11} \, \mathrm{erg \, s^{-1} \, cm^{-2}}$. Assuming a distance of 1.37~kpc \citep{Deller2012}, these flux levels transits to X-ray luminosities of $L_X = (2.2 \pm 0.2) \times 10^{33} \, \mathrm{erg \, s^{-1}}$ and $2.9^{+0.4}_{-0.2} \times 10^{33} \, \mathrm{erg \, s^{-1}}$. These values are in agreement with the sub-luminous disc state. Given the long-term stability of J1023 in this regime since 2013, it is highly probable that the source remained in this state also during the epoch of our GTC observations.

\section{Trailed spectra and Doppler maps} \label{appendix:B}

\begin{figure*}[h!]
\centering
{\includegraphics[width=.28\textwidth]{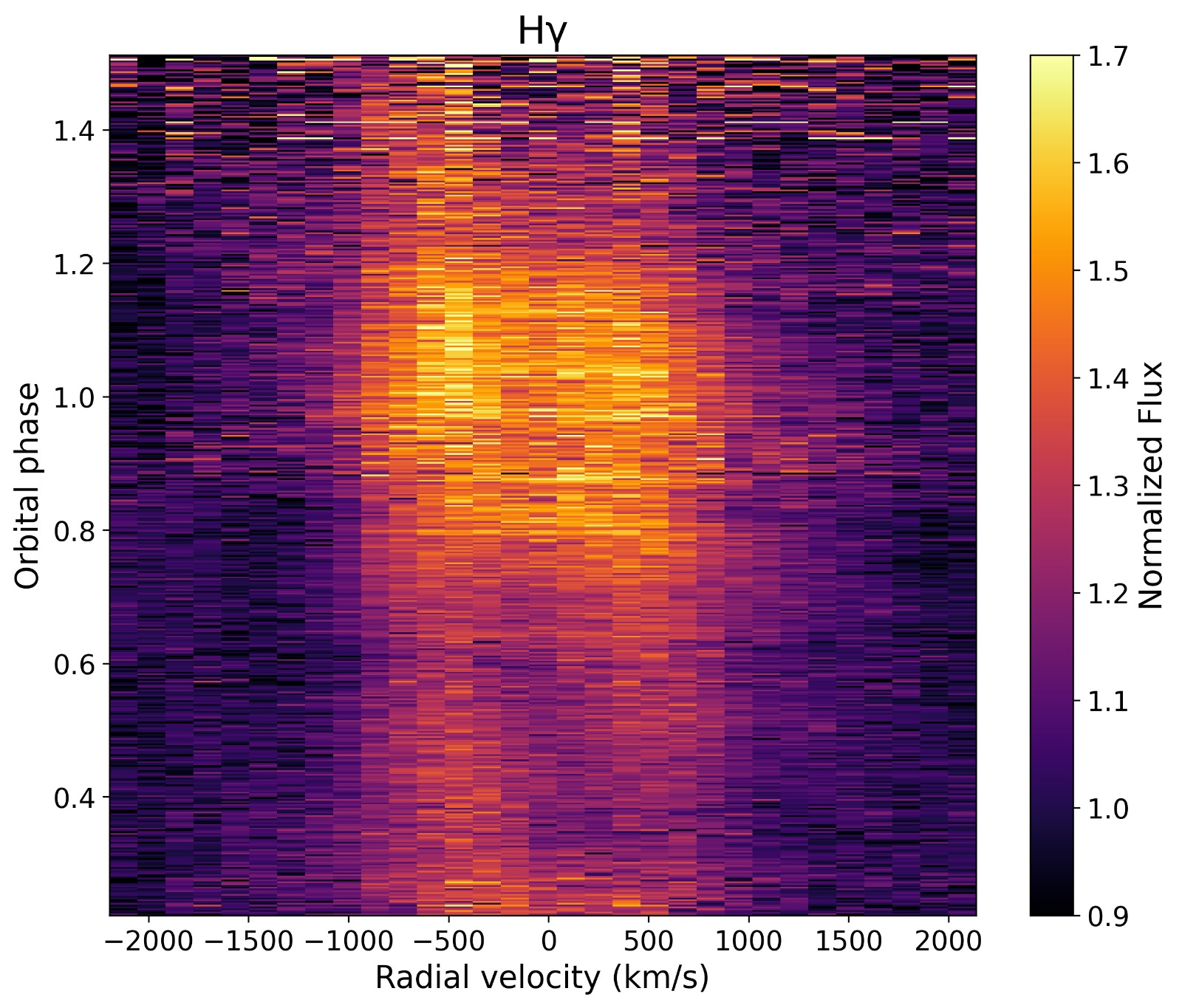}}\quad
{\includegraphics[width=.28\textwidth]{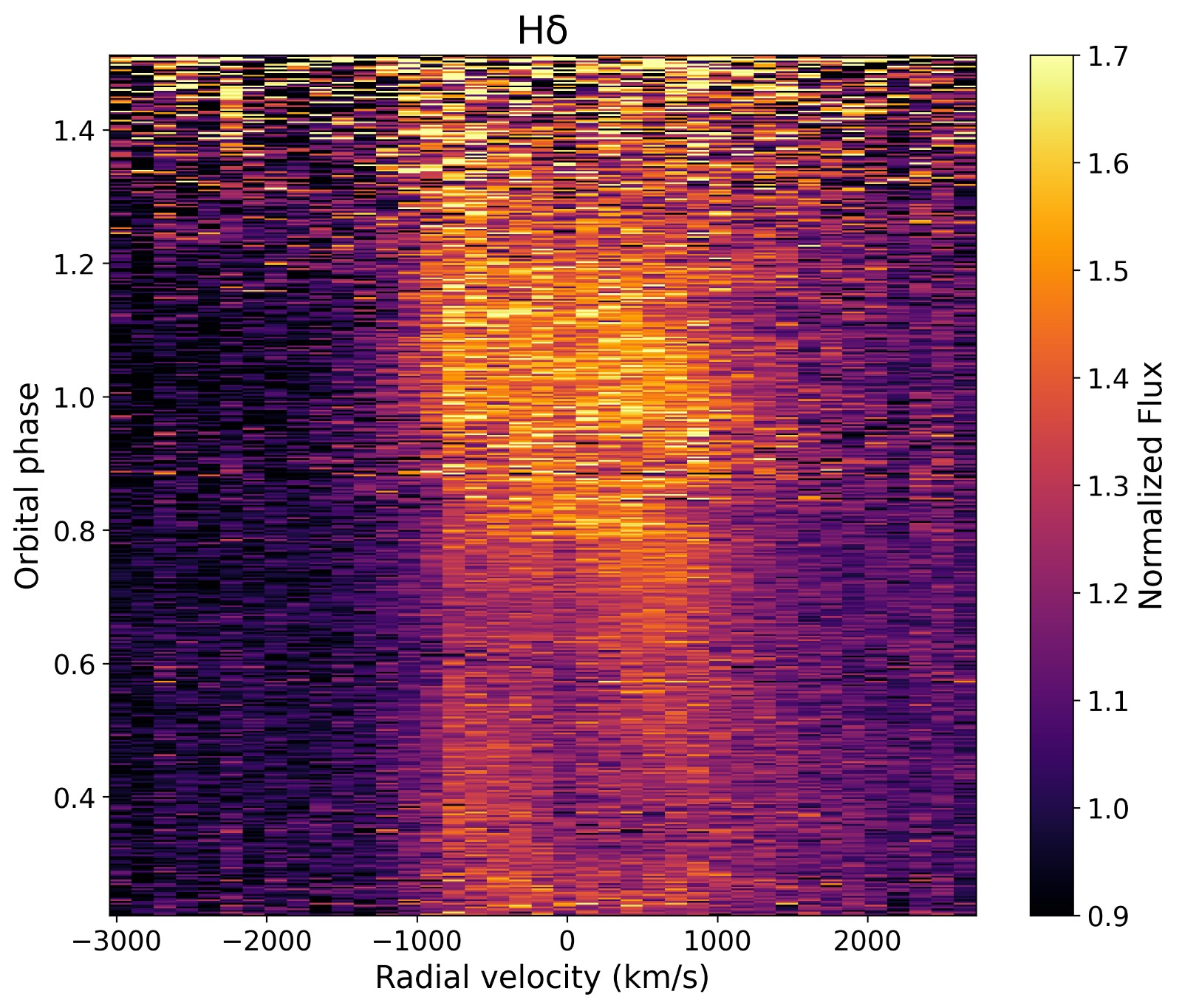}} \\
{\includegraphics[width=.28\textwidth]{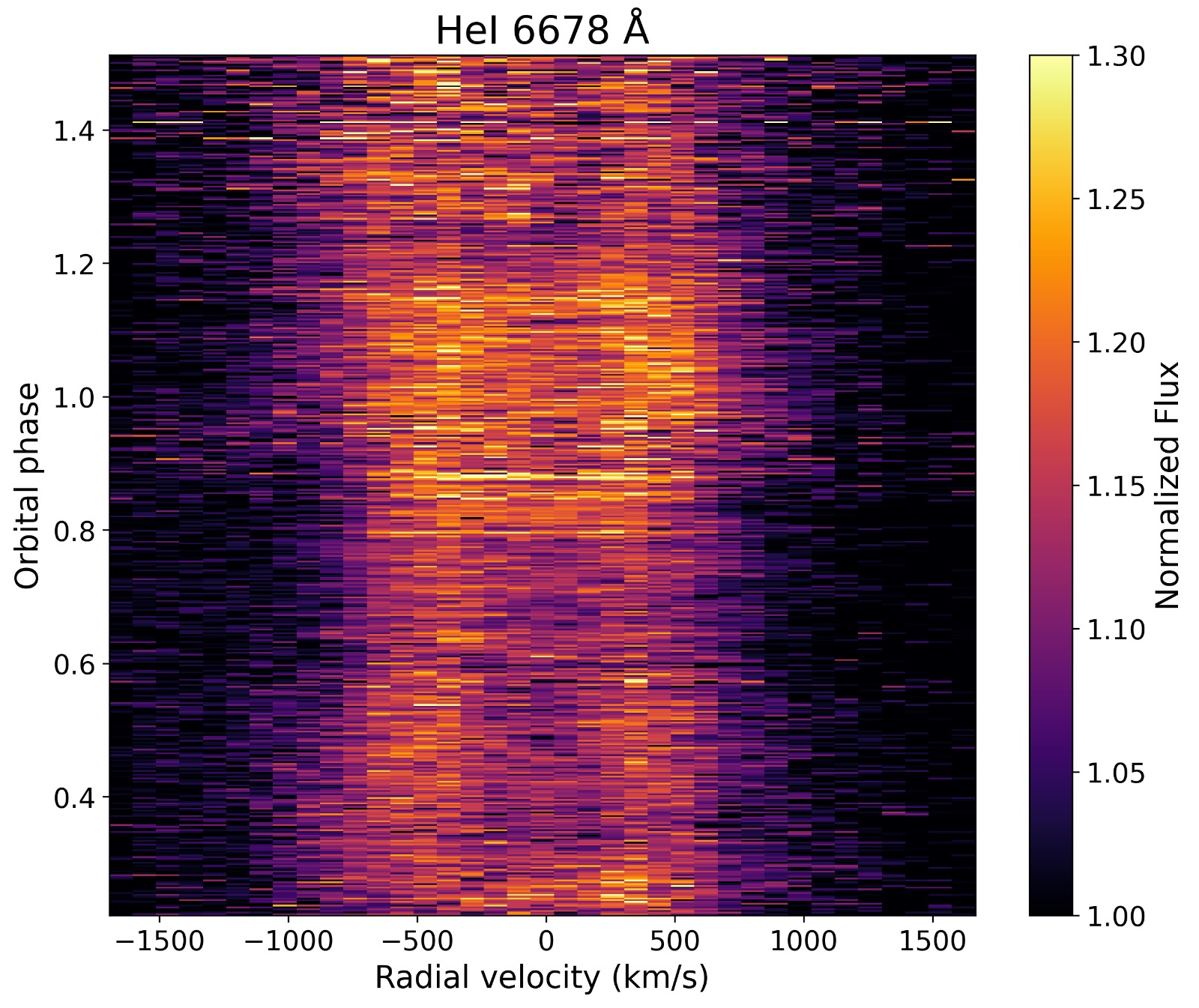}}\quad
{\includegraphics[width=.28\textwidth]{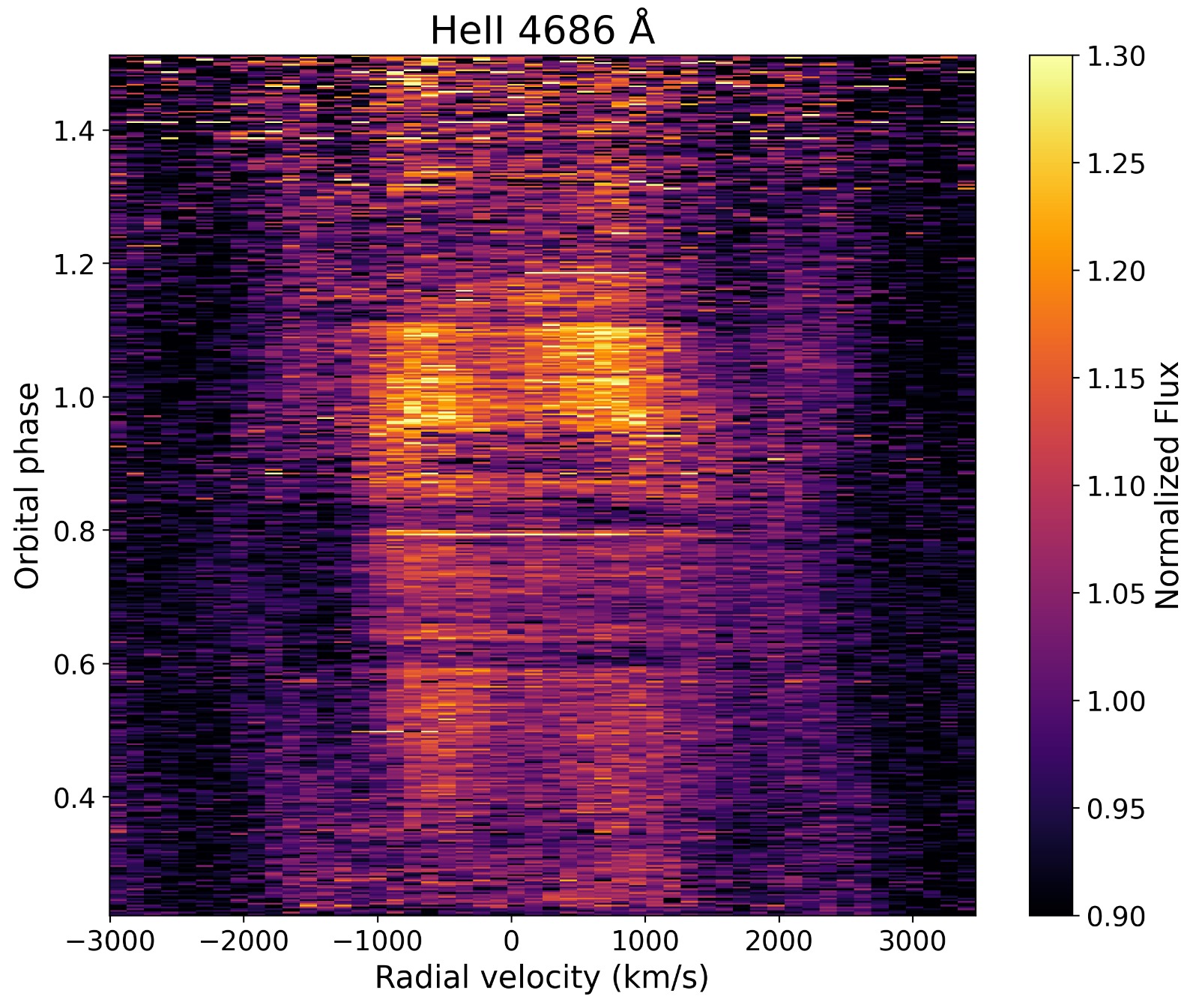}}\\
\caption{Trailed spectrograms for the H$\gamma$, H$\delta$, HeI at 6678 $\AA$ and HeII at 4686 $\AA$ emission lines.}
\label{fig:Trail_appendice}
\end{figure*}
\noindent

\begin{figure*}[h!]
\centering
{\includegraphics[width=.31\textwidth]{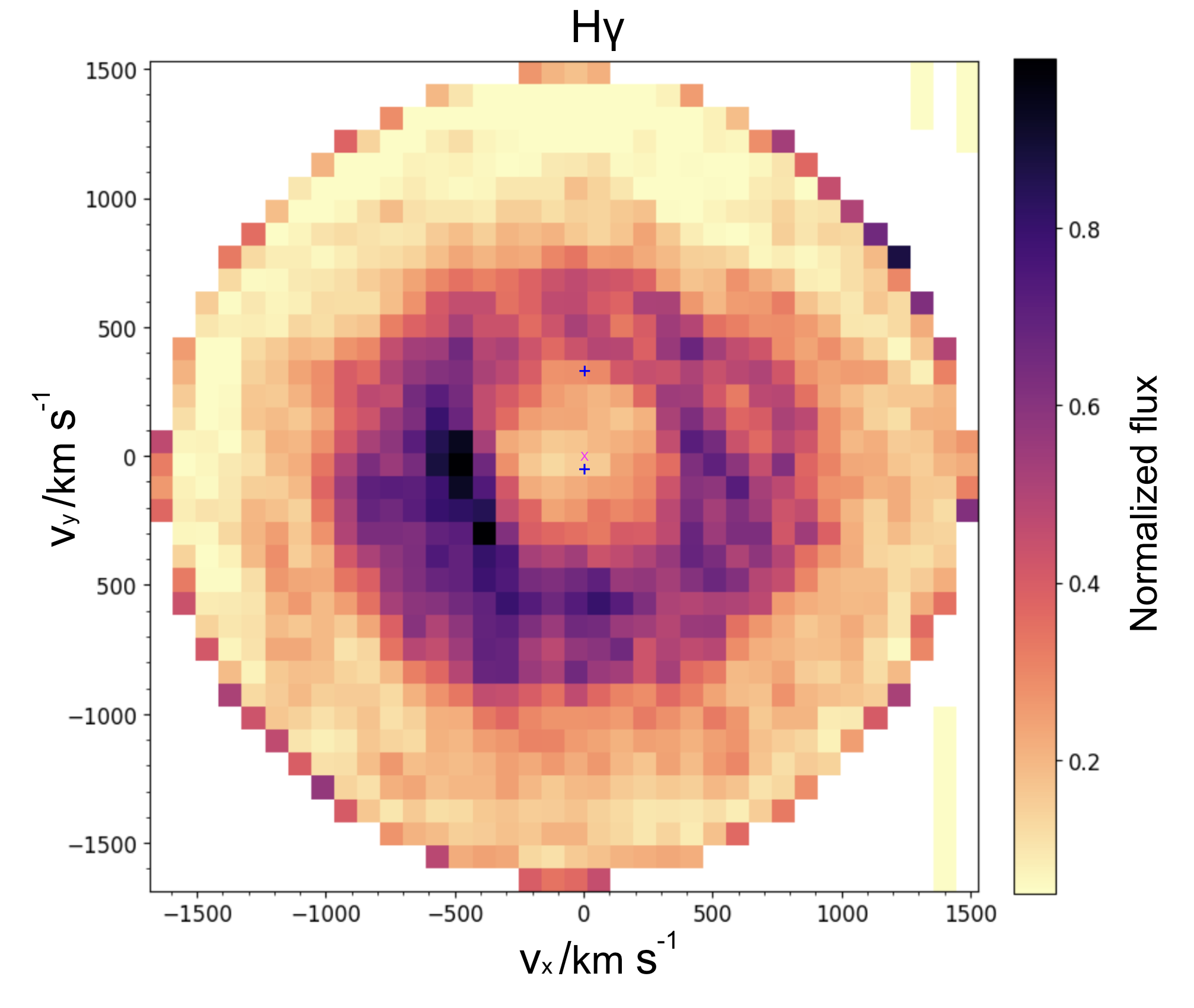}} \quad
{\includegraphics[width=.31\textwidth]{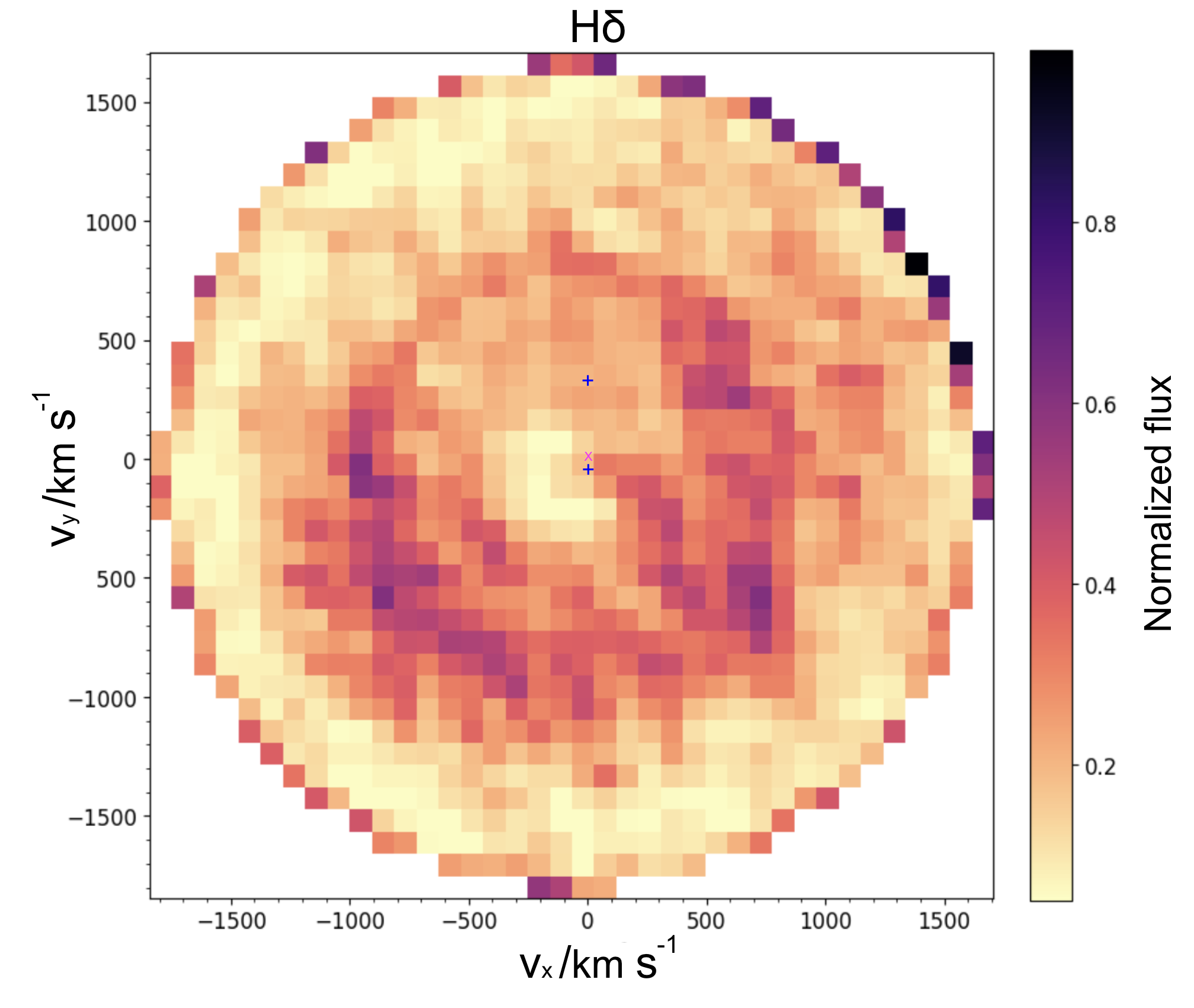}} \quad
{\includegraphics[width=.31\textwidth]{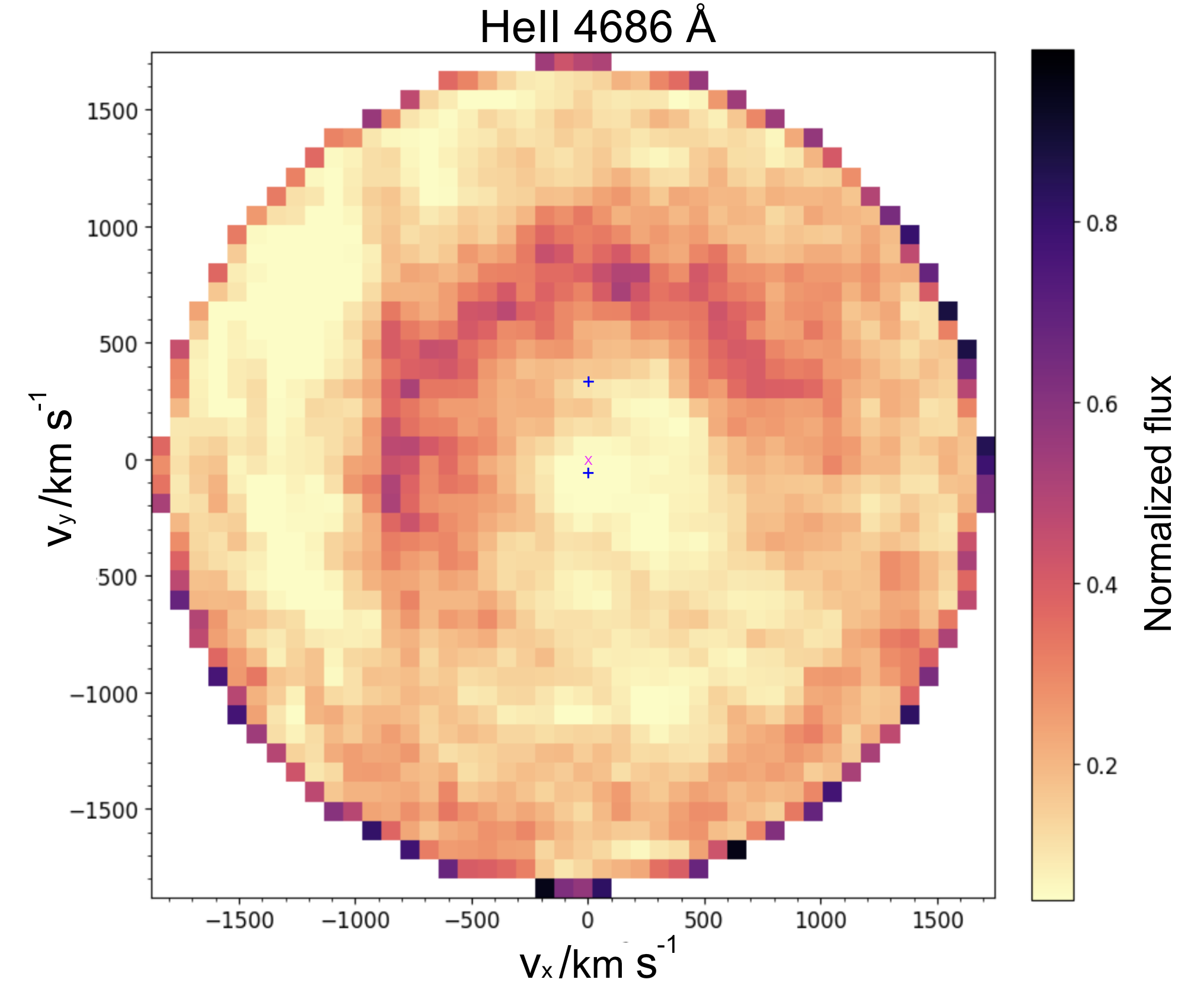}} \\
\caption{Doppler images in the velocity space for H$\gamma$, H$\delta$ and HeII at 4686 $\AA$ emission lines. The crosses indicate the positions of the compact object and the companion star, while the X marks the origin of the coordinate system.}
\label{fig:Doppler_appendice}
\end{figure*}
\noindent

\end{document}